\documentclass[prd,superscriptaddress,twocolumn,preprintnumbers,%
  amsmath,amssymb,longbibliography]{revtex4-1}
  
\usepackage[breaklinks,colorlinks=true,hidelinks]{hyperref}
\usepackage[T1]{fontenc}
\usepackage[utf8]{inputenc}
\usepackage{mathtools}
\usepackage{amsmath,amsthm,amssymb}
\usepackage{amsmath,bm}
\usepackage{bbold} %for fields and operators
\usepackage{dsfont}
\usepackage{lipsum}
\usepackage{calrsfs}
\DeclareMathAlphabet{\pazocal}{OMS}{zplm}{m}{n}

\usepackage{color}
\allowdisplaybreaks

\usepackage[dvipsnames]{xcolor}
\newcommand\myshade{85}
\colorlet{mylinkcolor}{BrickRed}
\colorlet{mycitecolor}{NavyBlue}
\colorlet{myurlcolor}{Aquamarine}
\hypersetup{
  linkcolor  = mylinkcolor!\myshade!black,
  citecolor  = mycitecolor!\myshade!black,
  urlcolor   = .,%myurlcolor!\myshade!black,
  colorlinks = true,
}

\newcommand{\rev}[1]{\textcolor{black}{#1}}

\newcommand{\dd}{\mathrm{d}}
\newcommand{\indep}{\perp \!\!\! \perp}
\begin{document}
% Iterations of the title
\title{Thermodynamics of exponential Kolmogorov-Nagumo averages}
\author{Pablo A. Morales}
\affiliation{Research Division, Araya Inc., 
Tokyo 107-6019, Japan}
\email{pablo$_$morales@araya.org}
\author{Jan Korbel}
\affiliation{Section for Science of Complex Systems, CeMSIIS, Medical University of Vienna, Spitalgasse, 23, 1090 Vienna, Austria}
\email{jan.korbel@meduniwien.ac.at}
\affiliation{Complexity Science Hub Vienna, Josefst\"{a}dter Strasse 39, 1080 Vienna, Austria}
\author{Fernando E. Rosas}
\affiliation{Department of Informatics, University of Sussex, Brighton BN1 9RH, UK}
\email{f.rosas@sussex.ac.uk} 
\affiliation{Centre for Psychedelic Research, Department of Brain Science, Imperial College London, London SW7 2DD, UK}
\affiliation{Centre for Complexity Science, Imperial College London, London SW7 2AZ, UK}
\affiliation{Centre for Eudaimonia and Human Flourishing, University of Oxford OX3 9BX, UK}

\newtheorem{definition}{Definition}
\newtheorem{theorem}{Theorem}
\newtheorem{lemma}{Lemma}
\newtheorem{proposition}{Proposition}
\newtheorem{corollary}{Corollary}
\newtheorem{example}{Example}
\newtheorem{remark}{Remark}

\begin{abstract}
This paper investigates generalized thermodynamic relationships in physical systems where relevant macroscopic variables are determined by the exponential Kolmogorov-Nagumo average.
We show that while the thermodynamic entropy of such systems is naturally described by R\'{e}nyi's entropy with parameter $\gamma$,
an ordinary Boltzmann distribution still describes their statistics under equilibrium thermodynamics.  
\rev{Our results show that systems described by exponential Kolmogorov-Nagumo averages can be interpreted as systems originally in thermal equilibrium with a heat reservoir with inverse temperature $\beta$ that are suddenly quenched to another heat reservoir with inverse temperature $\beta' = (1-\gamma)\beta$.} 
\rev{Furthermore, we show the connection with multifractal thermodynamics.} For the non-equilibrium case,
we show \rev{that} the dynamics of systems described by exponential Kolmogorov-Nagumo averages still observe a second law of thermodynamics and the H-theorem. 
We further discuss the applications of stochastic thermodynamics in those systems --- namely, the validity of fluctuation theorems --- and the connection with thermodynamic length. 
\end{abstract}

\maketitle

\section{Introduction}

In ancient Greece, three classic types of \rev{averages} were extensively studied: arithmetic mean $\frac{1}{n} \sum_i x_i$, geometric mean $(\prod_i x_i)^{1/n}$, and harmonic mean $n (\sum_i \tfrac{1}{x_i})^{-1}$, % These three types of averages --- collectively called \emph{Pythagorean means} --- 
which \rev{played various roles in} %naturally appeared 
physics, geometry, and music. 
These so-called \emph{Pythagorean means} found a natural generalisation via functional analysis and measure theory % the natural generalization of these  
into the well-known one-parametric class of \emph{H\"{o}lder means} $(\tfrac{1}{n}\sum_i x_i^p )^{1/p}$. 
%proved to be fertile for theoretical reasons while finding many applications in physics and elsewhere as well. 
\rev{A different generalization of the notion of average was independently proposed by Andrey Kolmogorov~\cite{Kolmogorov1930388} and Mitio Nagumo~\cite{Nagumo193071} in 1930}, which took the form of $f^{-1}\left(\frac{1}{n} \sum_i f(x_i) \right)$ for any continuous and injective function $f$. 
\rev{These \emph{Kolmogorov-Nagumo means} --- also known as quasi-arithmetic means or f-means --- 
%or . This broad class of averages 
have triggered numerous theoretical developments} by several researchers, including de Finneti~\cite{de1931sul}, Jessen~\cite{jessen}, Kitagawa~\cite{kitagawa1934some},  Acz\'{e}l~\cite{aczel1948mean}, or Fodor and Roubens~\cite{fodor1995meaningfulness}.

Kolmogorov-Nagumo averages have found applications in many fields, including machine learning~\cite{Nielsen16}, random fields~\cite{Porcu2009}, or fuzzy sets \cite{Klement99}. A particularly important application of Kolmogorov-Nagumo averages is the introduction of \emph{R\'{e}nyi entropy}~\cite{renyi1976}, which in turn have found many applications in quantum systems~\cite{stephan2014geometric,ShannonRenyiQuantumSpin,jizba2016one}, strongly coupled or entangled systems~\cite{Dong:2016fnf,Barrella:2013wja,jizba2019maximum}, phase transitions~\cite{GeometricMutInf,DetectingPhaseTwithRenyi,PhysRevLett.107.020402}, multifractal thermodynamics \cite{jizba2004,jizba04b}, and time series analysis~\cite{jizba2012renyi,jizba2014multifractal}, among others.
\rev{Another important application of generalized means have been %established in connection with 
the development of} the thermodynamics of complex systems~\cite{thurner2018}. \rev{In addition to concepts such as} deformed calculus \cite{borges2004,Czachor2020}, escort means \cite{beck1993,Abe03,Bercher11}, or non-linear dynamics~\cite{frank2004nonlinear,Korbel_2021}, Kolmogorov-Nagumo means served as a natural framework for generalized entropies \cite{czachor}
--- with the R\'{e}nyi entropy being a special case. %Special attention has been paid to the universality of L\'{e}gendre structure for the case of so-called generalized. 
\rev{Furthermore, connections between the maximum entropy principle and R\'{e}nyi entropy have been established} in Refs.~\cite{jizba2019maximum,PhysRevResearch.3.033216}

\rev{In this paper we investigate the thermodynamics of systems whose relevant macroscopic variables are determined by the exponential Kolmogorov-Nagumo average. 
The pioneering work of Czachor and Naudts~\cite{czachor} 
%The preliminary results discussing 
on the thermodynamics of exponential Kolmogorov-Nagumo averages discusses aspects of} equilibrium thermodynamics, 
%of exponential Kolmogorov-Nagumo averages, including the definition of R\'{e}nyi entropy and 
such as a generalized internal energy and its  %Furthermore, they discuss the 
relationship with 
%to the popular case of non-extensive 
Tsallis entropy~\cite{tsallis1988possible}. 
\rev{Bagci and Tirnakli~\cite{BAGCI20093230} 
%follow up by discussing 
studied a generalized maximum entropy principle for the exponential Kolmogorov-Nagumo averages, while leaving its thermodynamic interpretation unclear}. A first thermodynamic interpretation of R\'{e}nyi entropy was given by Baez \cite{baez}, who noted that the R\'{e}nyi entropy of a Boltzmann distribution is equal to the change of Helmholtz free energy divided by the change of temperature. \rev{The first goal of this paper is to develop a framework that unifies and extends these previous results}, enabling a unified interpretation of all thermodynamic quantities as well as the maximum entropy principle in such scenarios.

A second set of relevant results is related to the L\'{e}gendre structure of thermodynamics \cite{plastino1997}.
Scarfone, Matsuzoe and Wada~\cite{scarfone2022} showed that the L\'{e}gendre structure of equilibrium thermodynamics remains valid for  Kolmogorov-Nagumo averages on the level of macroscopic quantities. Additionally, Wong \cite{wong2018logarithmic,wong2019logarithmic} introduced a generalized L\'{e}gendre structure that corresponds to the R\'{e}nyi entropy. 
%leading to generalized divergence of Bregman type. The generalized Bregman divergence is based on the generalization of the \emph{link function} \cite{wong2019logarithmic}. 
\rev{A second goal of this paper clarify the thermodynamic interpretation of this generalized L\'{e}gendre structure.}
%is to show that while the ordinary L\'{e}gendre structure corresponds to the equilibrium free energy, the generalized L\'{e}gendre structure corresponds to free energy difference measuring the reversible work of a process starting in equilibrium and ending in an arbitrary state. 
Furthermore, Liangrong, Hong, and Liu \cite{PhysRevE.101.022114} have recently discussed non-equilibrium thermodynamics with non-extensive quantities. 
Investigations on non-equilibrium thermodynamics and multifractals related to dielectric breakdown was carried out by Enciso and colleages~\cite{Enciso_2019}. 
We will show that these results are closely related to non-equilibrium stochastic thermodynamics based on exponential Kolmogorov-Nagumo averages, \rev{while deriving a non-equilibrium version of the second law of thermodynamics, 
%definition of thermodynamic forces and fluxes, 
a generalized H-theorem, 
%as with the H function being expressed in terms of R\'{e}nyi-Bregmann divergence, the explicit formula for the entropy production in terms of R\'{e}nyi-Csisz\'{a}r divergence 
and proving the validity of detailed and integrated fluctuation theorem.} 
Finally, several authors have studied generalized statistical mechanics from the point of information geometry \cite{amari2016information,naudts2008generalised,korbel2019information,korbel2020information}. In particular, Eguchi, Komori, and Ohara have investigated the geometry of generalized e-geodesic and m-geodesic in terms of Kolmogorov-Nagumo means \cite{eguchi2016information}. We extend their results to the case of a generalized %Bregman 
divergence, and calculate relevant quantities such as Fisher-Rao information and thermodynamic length. 

The rest of the paper is organized as follows. Section~\ref{sec:thermo} defines the exponential Kolmogorov-Nagumo means and summarizes its main properties. 
%as well as connections with cumulants, large deviation theory, and the R\'{e}nyi entropy. 
Section~\ref{sec:Kol_nag} then establishes equilibrium thermodynamics based on Kolmogorov-Nagumo averages of both entropy and internal energy. 
%, presenting a calculation of the maximum-entropy distribution and discussing its thermodynamic interpretation. 
\rev{Section~\ref{sec:multifractals} describes the application of this framework to multifractal systems.} 
%, and how the application of Kolmogorov-Nagumo averages establishes a connection between two distinct multifractal formalisms.}
Section~\ref{sec:non-equi} is focused to the discussion on non-equilibrium thermodynamics. 
%We start with the L\'{e}gendre structure of thermodynamics of the free energy difference between equilibrium and non-equilibrium states, which can be expressed in terms of R\'{e}nyi-Bregmann divergence, and leads towards a novel formulation of the H-theorem for exponential Kolmogorov-Nagumo averages. The section further focuses on extending stochastic thermodynamics to Kolmogorov-Nagumo averages, mainly the second law of thermodynamics, thermodynamic forces and fluxes, entropy production, and fluctuation theorems. 
Section~\ref{sec:thermo_length} is then focused on thermodynamic length. 
Finally, Section~\ref{sec:conclusion} summarizes our main conclusions. %The paper is also equipped with two appendices, showing finer technical details of some calculations.  

\section{Exponential Kolmogorov-Nagumo averages}
\label{sec:thermo}

%*** TODO ***
%In thermodynamics and statistical mechanics the L\'{e}gendre transform plays a crucial role in establishing a bridge between intensive and extensive variables. 
%--- where often one is easier to measure and control than its conjugate counterpart. 

Our line of \rev{inquiry} focuses on systems governed by constraints that can be expressed in terms of non-linear averages. A natural extension of linear (arithmetic) averages is the \textit{Kolmogorov-Nagumo average}~\cite{Kolmogorov1930388,Nagumo193071} 
\begin{equation}
\langle X \rangle_f = f^{-1}\left(\sum_i p_i f(x_i)\right),    
\end{equation}
where $f$ is a continuous injective function. Without loss of generality, we focus on cases where $f$ is an increasing function. Note that the average is invariant to affine transformations of the function $f(x) \mapsto f_{a,b}(x) = a f(x) + b$, i.e., $\langle X \rangle_f = \langle X \rangle_{f_{a,b}}$.

The linear average is recovered by setting $f(x)=x$, which can be shown to be the only one that satisfies two properties:
\begin{enumerate}
    \item \emph{Homogeneity:} $\langle a X \rangle = a \langle X \rangle$\, ,
    \item \emph{Translation invariance:} $\langle X + c \rangle = \langle X \rangle + c$\, .
\end{enumerate}
In \cite{aczel1989functional}, it is shown that the first condition alone leads to functions of the form $f(x) = x^p$, which corresponds to the well-known class of H\"{o}lder averages. In contrast, the second property has been shown to lead to the class of so-called \emph{exponential Kolmogorov-Nagumo averages}~\cite{hardy1952inequalities}, which corresponds to $f(x) = \exp_{\gamma}(x) = (e^{\gamma x}-1)/\gamma$ and inverse function $\ln_\gamma(x) = \tfrac{1}{\gamma}\ln(1+\gamma x)$. 

The property of translation invariance is of particular interest for statistical mechanics, as it guarantees that thermodynamic relations are independent of the specific value of the \rev{ground state energy}. % (i.e., one can always subtract the ground energy value). 
Also, the standard arithmetic mean is recovered for $\gamma = 0$. 
%
%Hence, this section focuses on developing the thermodynamic implications of this second class of averages.
%While thermodynamics of Kolmogorov-Nagumo averages has been discussed in Refs.~\cite{czachor,scarfone2022}, the connection between the exponential Kolmogorov-Nagumo means and generalized L\'{e}gendre transforms has not been explored yet. 
%
These considerations make us focus on the Kolmogorov-Nagumo averages, which lead to the following type of averages:
\begin{equation}
\langle X \rangle_\gamma = \ln_\gamma \left(\sum_i p_i \exp_\gamma(x_i)\right) = \frac{1}{\gamma}\ln \left(\sum_i p_i e^{\gamma x_i}\right).
\end{equation}
A key property of the exponential mean is a weaker version of the additivity of the expectation value, which reads
\begin{equation}
\langle X + Y \rangle_\gamma = \langle X \rangle_\gamma + \langle Y \rangle_\gamma \Leftarrow X \indep Y     ,
\end{equation}
where $\indep$ denotes statistical independence. In the general case, it is possible to express the expected value of sums as
\begin{eqnarray}\label{eq:cond}
\langle X + Y \rangle_\gamma &=& 
%\frac{1}{\gamma} \ln \sum_{ij} p_{ij} e^{\gamma(x_i+y_j)}  \nonumber\\
%&=& \frac{1}{\gamma} \ln \sum_i p_i e^{\gamma x_i} \sum_j  p_{j|i} e^{\gamma y_j} \nonumber\\
%&=& \frac{1}{\gamma} \ln \sum_i p_i e^{\gamma (x_i + \langle Y|X = X_i \rangle_\gamma)} \nonumber\\
%&=&
\langle X + \langle Y|X \rangle_\gamma \rangle_\gamma\, ,
\end{eqnarray}
where $\langle Y|X \rangle_\gamma$ is the conditional average, i.e.,
\begin{equation}
\langle Y|X \rangle_\gamma = \frac{1}{\gamma} \ln \sum_i p_i \, \exp( \gamma \langle Y|X=x_i\rangle_\gamma)~.
\end{equation}
¨
\subsection*{Connection with cumulants}
The exponential Kolmogorov-Nagumo averages are closely related to the cumulant-generating functions, which are given by
\begin{equation}\label{eq:cum-gen}
M_\gamma(X) = \ln \langle e^{\gamma X} \rangle = \gamma \cdot \langle X \rangle_\gamma    .
\end{equation}
By considering the well-known Taylor expansion of the cumulant-generating function, one can find that
\begin{equation}\label{eq:cumulants}
\langle X \rangle_\gamma = \sum_{n=1}^\infty \kappa_n(X) \frac{\gamma^{n-1}}{n!}    ,
\end{equation}
where $\kappa_n(X)$ is the $n$-th cumulant. This shows that the exponential Kolmogorov-Nagumo averages combine all cumulants weighted by the factor $\gamma^{n-1}/n!$. 

The relationship between cumulants and  Kolmogorov-Nagumo averages can be used to provide a complementary view of the properties of the latter. For example, 
the additivity property of the Kolmogorov-Nagumo average can be understood as a consequence of the additivity of the cumulant generating function for independent random variables. 
%In \rev{fact}, using Eq.~\eqref{eq:cumulants} one can find that
%\begin{eqnarray}
%\langle X + Y \rangle_\gamma &=&  \sum_{n=1}^\infty \frac{\gamma^{n-1}}{n!} \sum_{j=0}^n \binom{n}{j} \kappa_n(\underbrace{X,\dots,X}_j,\underbrace{Y,\dots,Y}_{n-j})\nonumber\\
%&=& \langle X \rangle + \langle Y \rangle \nonumber\\
%&&+ \frac{\gamma}{2} (\mathrm{Var}(X) + 2 \mathrm{Cov}(X,Y) + \mathrm{Var}(Y)) \nonumber\\
%&&+ \mathcal{O}(\gamma^2)~.
%\end{eqnarray}
%where $\kappa_n$ is the $n$-th multivariate moment. 
%Finally, Eq.~\eqref{eq:cond} can be obtained via applying the Law of total cumulance~\cite{totalcumulance}.

\subsection*{Connections with large deviation theory}
When considering the sum $S_n = \frac{1}{n} \sum_{i=1}^n X_i$ of $n$ i.i.d. random variables $X_1,\dots,X_n$, large deviation theory~\cite{touchette2009large} states that the probability of observing $S_n = s$ can be expressed \rev{as
%\begin{equation}
%P(S_n = s) \asymp e^{- n I(s)}~,
%\end{equation}
%\rev{where
\begin{equation}
\lim_{n \rightarrow \infty} \frac{1}{n} \log P(S_n=s) = I(s)  
\end{equation}
with $I(s)$ corresponding to the} so-called rate function. 
Cram\'{e}r's theorem states that the rate function can be obtained from the cumulant-generating function~\cite{touchette2009large}, which --- via Eq.~\eqref{eq:cum-gen} --- can be expressed in terms of $\gamma$-average using $\gamma=nk$ as follows:
\begin{equation}\label{eq:rate_func}
I(s) = \sup_{k} k \left(s - \langle S_n \rangle_{n k}\right)    
\end{equation}
where the $nk$-average can be expressed as
\begin{equation}
\langle S_n \rangle_{nk} = \frac{1}{nk} \ln \int \mathrm{d} s\,  e^{n k s} P(S_n = s)~.
\end{equation}
A connection between large deviation theory and statistical mechanics can then be drawn using Eq.~\eqref{eq:rate_func} by considering $k=-\beta$ being the inverse temperature and $X_i = h_i$ being the energy of $i$-th subsystem. Then, one can find that 
\begin{equation}
\langle S_n \rangle_\beta = -\frac{1}{\beta n} \langle e^{-\beta n h_n} \rangle = \Psi(\beta)~,
\end{equation}
with $\Psi = S - \beta U$ being the Free entropy (also known as Massieu function), with $S$ being the thermodynamic entropy and $U$ being the internal energy. For more details about this relationship, we refer the interested reader to Ref.~\cite{touchette2009large}.

\subsection*{R\'{e}nyi entropy as the exponential Kolmogorov-Nagumo average}
The R\'{e}nyi entropy can be naturally formulated in terms of the exponential Kolmogorov-Nagumo average of Hartley information $\ln 1/p_k$ \cite{hartley1928}, also known as the Shannon pointwise entropy~\cite{ince2017measuring}. To show this, let us consider $X$ to be a random variable with probability distribution $p_k$ and calculate the following:
\begin{eqnarray}\label{eq:renyi_entropy}
\frac{1}{1-\gamma}\left \langle \ln \frac{1}{p_k} \right \rangle_\gamma
&=& \frac{1}{1-\gamma} \ln_\gamma \left(\sum_i p_i e_\gamma(\ln 1/p_i) \right)\nonumber\\ 
&=& \frac{1}{\gamma(1-\gamma)} \ln \sum_{i} p_i^{1-\gamma}\\
&:=& R_\gamma(X)\,,
\end{eqnarray}
where $R_\gamma(X)$ denotes the R\'{e}nyi entropy of order $1-\gamma$. This result shouldn't be surprising since the R\'{e}nyi entropy appears in the Campbell coding theorem as the minimal price that one must pay to encode a message, where the price is the exponential function of the message length~\cite{campbell65}.

Note that the definition of the R\'{e}nyi entropy that we are using includes a factor $1/(1-\gamma)$ that is often not considered. We include this factor in this paper as its addition greatly simplifies the calculations presented in the next sections, which in turn will endow it with a clear thermodynamic meaning. 
Additionally, by including this factor the limit $\gamma \rightarrow 1$ of $R_\gamma(X)$ leads to the well-known Burg entropy $R_1 = - \sum_i \ln p_i$~\cite{Burg}. 

Let us finalize this section by noting that the joint R\'{e}nyi entropy can be decomposed as follows:
\begin{eqnarray}
R_\gamma(X,Y) &=& \frac{1}{1-\gamma} \left\langle \ln \frac{1}{p_{ij}} \right\rangle_\gamma \nonumber\\
&=& \frac{1}{\gamma} \left\langle \ln \frac{1}{p_i} + \left\langle \ln \frac{1}{p_{j|i}} \Bigg| p_i \right\rangle_\gamma \right\rangle_\gamma\nonumber\\
&=& \frac{1}{\gamma(1-\gamma)} \ln \sum_i p_i^{1-\gamma}\nonumber\\
&&+ \sum_i p_i \frac{1}{\gamma(1-\gamma)} \ln \sum_j p_{j|i}^{1-\gamma}   \nonumber\\
&=& R_\gamma(X) + R_\gamma(Y|X)~,
\end{eqnarray}
where 
%the first term is the R\'{e}nyi entropy of $X$ and the second term 
$R_\gamma(Y|X)$ is known as the conditional R\'{e}nyi entropy.

\section{Equilibrium thermodynamics} \label{sec:Kol_nag}

Let us now consider a system whose internal energy at state $i$ is given by $\epsilon_i$. Then, the Kolmogorov-Nagumo $\gamma$-average energy of the system is calculated as $U = \langle \epsilon \rangle_\gamma = \ln \sum_i p_i \exp(\epsilon_i)$. However, as discussed in the previous section, this quantity is not invariant to rescaling by a factor (i.e., $\epsilon_i \mapsto \lambda \epsilon_i$), and furthermore, it  does not have properly defined units of energy. 
Therefore, it will be convenient to focus our analysis on the following rescaled internal energy:
\begin{equation}
 U^\beta_\gamma := \frac{1}{\beta}\langle \beta \epsilon \rangle_\gamma = \frac{1}{\beta} \ln_\gamma \left(\sum_i p_i e_\gamma(\beta \epsilon_i)\right),   
\end{equation}
where $\beta$ is the inverse temperature of the system. 
Note that the units of $\beta$ are 1/Joules, and hence $\beta \epsilon_i$ is dimensionless, making $U^\beta_\gamma$ a properly defined mean energy. This type of average energy function has already been considered in previous research, e.g., in Ref.~\cite{czachor}. A summary of thermodynamic quantities studied in this work is presented in Table~\ref{tab:my_label}
\begin{table*}[t]
    \centering
    \begin{tabular}{|l|l|l|}
\hline
    Quantity & Shannon thermodynamics & R\'{e}nyi thermodynamics
    \\
\hline
Entropy & $S(p) = - \sum_k p_k \ln p_k$    & $R_\gamma(p) = \frac{1}{\gamma(1-\gamma)} \ln \sum_k p_k^{1-\gamma}$\\
Cross-Entropy & $S(p,q) = - \sum_k p_k \log q_k$ & $ R_\gamma(p,q) = - \frac{1}{\gamma} \ln \sum_k p_k q_k^{-\gamma} - \gamma R_\gamma(q)$ \\
Csisz\'{a}r Divergence & $\mathcal{D}_\text{KL}(p||q) = \sum_k p_k \ln(p_k/q_k)$ & $\mathfrak{D}_\gamma(p||q) =  \frac{1}{\gamma} \ln \sum_k p_k^{\gamma+1}q_k^{-\gamma}$\\
Bregmann Divergence & $\mathcal{D}_\text{KL}(p||q) =  - S(p) - S(p,q)$ &
    $\mathcal{D}_\gamma(p||q) = -R_\gamma(p) - R_\gamma(p,q)$ \\
Link function & $C(x,y) = \sum_k x_k y_k$ & $C_\gamma(x,y) = \frac{1}{\gamma} \ln(1+\gamma \sum_k x_k y_k)$\\ 
Kolmogorov-Nagumo function & $g(x) = x =  g^{-1}(x)$ & $g(x) = \frac{e^{\gamma x}-1}{\gamma}$, \quad $g^{-1}(x) = \frac{1}{\gamma}\ln(1+\gamma x)$\\
Internal energy & $U = \sum_k p_k \epsilon_k$ & $U_\gamma = \frac{1}{\gamma} \ln \left(\sum_k p_k e^{\gamma \epsilon_k}\right)$
\\
Heat rate & $\dot Q = \sum_k \dot{p}_k \epsilon_k$ & $\dot{Q}^\beta_\gamma = \frac{1}{\beta \gamma}\frac{\sum_k \dot{p}_k e^{\gamma \beta \epsilon_k}}{\sum_k p_k e^{\gamma \beta \epsilon_k}}$
\\
Work rate & $\dot{W} = \sum_k p_k \dot{\epsilon}_k$ & $\dot{W}^\beta_\gamma = \frac{\sum_k  p_k \dot{\epsilon}_k e^{\gamma \beta \epsilon_k}}{\sum_k p_k e^{\gamma \beta \epsilon_k}}$\\
Equilibrium free energy & $F(\pi) = -\frac{1}{\beta} \ln \sum_k e^{-\beta \epsilon_k}$ & $F_\gamma(\pi) = \frac{1}{\beta(\gamma-1)} \ln \sum_k e^{(\gamma-1)\beta \epsilon_k}$\\ 
Non-equilibrium free energy &
$F(p) = F(\pi) + \frac{1}{\beta} \mathcal{D}_\text{KL}(p||\pi)$ & $F_\gamma(p) = F_\gamma(\pi) + \frac{1}{\beta} \mathcal{D}_\gamma(p||\pi)$\\
Entropy production & $\Delta \Sigma = \mathcal{D}_{KL}(P(x(t))||\tilde{P}(\tilde{x}(t)))$ & $\Delta\Sigma_\gamma = \mathfrak{D}_{\gamma}(P(x(t))||\tilde{P}(\tilde{x}(t)))$\\
\hline
    \end{tabular}
    \caption{Comparison of the relevant quantities for Shannon's and R\'{e}nyi's framework. Quantities from standard the Shannon framework can be recovered by taking $\gamma \rightarrow 0$.}
    \label{tab:my_label}
\end{table*}

Similarly, our \rev{analysis} will consider the thermodynamic entropy as defined by the Kolmogorov-Nagumo average of $\ln 1/p_k$, which gives us R\'{e}nyi entropy, as discussed in the previous section. 
Interestingly, for the case $\gamma=0$ this formalism recovers the standard definitions of average energy and Boltzmann-Gibbs entropy, while for $\gamma\neq1$ it accounts for different scenarios --- which we study in the \rev{following}.

\subsection*{Maximum entropy principle}
Let us now focus on \rev{distributions} that \rev{correspond} to a given value of mean energy as given by $U_\gamma^\beta$, according to the maximum entropy principle. The distribution $\pi_i$ that maximizes the R\'{e}nyi entropy while satisfying a given $\gamma$-average level of energy can be found by using the method of Lagrange multipliers on the following Lagrange function:
\begin{equation}
    \mathcal{L} = R_\gamma - \alpha_0 \sum_i p_i - \alpha_1 \frac{1}{\beta}\langle \beta \epsilon \rangle_\gamma~. 
\end{equation}
A direct calculation shows that $\pi_i$ is the solution of the following equation:
\begin{equation}
\frac{1}{\gamma} \frac{\pi_i^{-\gamma}}{\sum_k \pi_k^{1-\gamma}} - \alpha_0 - \frac{\alpha_1}{\beta \gamma} \frac{e^{\gamma \beta \epsilon_i}}{\sum_k \pi_k e^{\gamma \beta \epsilon_k}} = 0~.
\end{equation}
By multiplying the equation by $\pi_i$ and summing over $i$, one obtains that $\alpha_0 = \frac{1-\alpha_1}{\gamma}$, which leads to 
\begin{equation}
\frac{\pi_i^{-\gamma}}{\sum_k \pi_k^{1-\gamma}}-1 = \frac{\alpha_1}{\beta} \left( \frac{e^{\gamma\beta \epsilon_i}}{\sum_k \pi_k e^{\gamma\beta \epsilon_k}}-1 \right).
\end{equation}
Above, the Lagrange parameter $\alpha_1$ 
can be chosen such that one recovers standard thermodynamic relationships. To this end, we identify that $\alpha_1=\beta$ (which is the standard relation between the Lagrange multiplier and inverse temperature), which gives us
\begin{equation}\label{eq:equi}
\pi_i =  \frac{\left(\sum_k \pi_k e^{\gamma \beta \epsilon_k}\right)^{1/\gamma}}{\left(\sum_k \pi_k^{1-\gamma}\right)^{1/\gamma}}  \exp(-\beta \epsilon_i) = \frac{\exp(-\beta \epsilon_i)}{Z^\beta}\,
\end{equation}
which is just the Boltzmann distribution with inverse temperature $\beta$. We can rewrite the previous relationship as
\begin{equation}
\ln \pi_i = - \beta(\epsilon_i - U_\gamma^\beta(\pi)) - (1-\gamma) R_\gamma(\pi)\,    
\end{equation}
and using the fact that $\sum_k \pi_k = 1$ one finds that
\begin{eqnarray}
\ln \sum_k e^{-\beta \epsilon_k} &=& \frac{1}{\gamma} \ln \sum_k \pi_k^{1-\gamma} - \frac{1}{\gamma} \ln \sum_k \pi_k e^{\gamma \beta \epsilon_k}\nonumber\\
&=&(1-\gamma) R_\gamma - \beta U^\beta_\gamma \nonumber\\
&=& \Psi_\gamma - \gamma R_\gamma\, ,\label{eq:cool_factor}
\end{eqnarray}
where $\Psi_\gamma = R_\gamma - \beta U_\gamma^\beta$ is the free entropy (also called Massieu function). Finally, one can derive the free energy by plugging in the  equilibrium distribution into internal energy and R\'{e}nyi entropy as
\begin{equation}
\label{eq:f2}
F^\beta_\gamma(\pi) = U_\gamma^\beta(\pi)- \frac{1}{\beta} R_\gamma(\pi) = \frac{1}{(\gamma-1) \beta} \ln \sum_k e^{(\gamma-1)\beta \epsilon_k}.
\end{equation}
Thus, one finds that $F_\gamma^\beta(\pi) = F^{\beta(1-\gamma)}(\pi)$. This result shows that the Kolmogorov-Nagumo average is effectively a rescaling of the free energy from inverse temperature $\beta$ to $(1-\gamma)\beta$.  Note that Eq.~\eqref{eq:f2} recapitulates the standard relationship between the free energy and the partition function for $\gamma=0$.

These results reveal that, perhaps surprisingly, the obtained equilibrium distribution in Eq.~\eqref{eq:equi} (obtained via the maximum entropy principle) is Boltzmann, while the thermodynamic quantities at play are nonetheless different from the case of ordinary thermodynamics based on Shannon entropy and linear averages. In fact, Eq.~\eqref{eq:cool_factor} implies that the free entropy and the logarithm of the partition function generally are not equal but differ by the term $-\gamma R_\gamma$, which vanishes only for $\gamma=0$.

\subsection*{Thermodynamic interpretation}
Let us now focus on the thermodynamic interpretation of equilibrium thermodynamics with the exponential Kolmogorov-Nagumo average. We denote the equilibrium versions of thermodynamic potentials by calligraphic symbols, i.e., 
\begin{eqnarray}
\mathcal{U}_\gamma^\beta &=& U_\gamma^\beta(\pi)~,\\       
\mathcal{R}_\gamma^\beta &=& R_\gamma(\pi)~, \label{eq:Renyi_eq}\\
\mathcal{F}_\gamma^\beta &=& \mathcal{U}_\gamma^\beta - \frac{1}{\beta} \mathcal{R}_\gamma^\beta~.
\end{eqnarray}
Let us start with Eq.~\eqref{eq:Renyi_eq}. As already shown in Ref.~\cite{baez}, the equilibrium R\'{e}nyi entropy can be expressed as
\begin{eqnarray}
\mathcal{R}_\gamma^\beta &=& \frac{1}{\gamma(1-\gamma)} \ln \sum_i \left(\frac{e^{-\beta \epsilon_i}}{Z^\beta}\right)^{1-\gamma}\nonumber\\
&=& \frac{1}{\gamma(1-\gamma)} \ln \sum_i e^{-(1-\gamma)\beta} - \frac{1}{\gamma} \ln Z^\beta\nonumber\\
&=& -\frac{\beta}{\gamma} \left(\mathcal{F}^{(1-\gamma) \beta} -  \mathcal{F}^{\beta}\right).
\end{eqnarray}
By defining 
\begin{equation}
\beta' = (1-\gamma)\beta  \quad \Rightarrow \quad \gamma = 1-\frac{\beta'}{\beta}~,
\end{equation}
we obtain 
\begin{equation}
 \mathcal{R}_\gamma^\beta =  \beta^2 \, \frac{\mathcal{F}^{\beta'} - \mathcal{F}^\beta}{\beta'-\beta}~,  
\end{equation}
which is the $\beta$ rescaling of the free energy difference. This can be interpreted as the maximum amount of work the system can perform by quenching the system from inverse temperature $\beta$ to inverse temperature $\beta'$. Note that $\gamma \rightarrow 0$ corresponds to $\beta' \rightarrow \beta$ and we recover the ordinary relation between entropy and free energy
\begin{equation}
\mathcal{S}^\beta = \beta^2 \left(\frac{\partial \mathcal{F}^\beta}{\partial \beta}\right)~.
\end{equation}
where $\mathcal{S}^\beta$ is the ordinary thermodynamic entropy. 

Since $ \mathcal{F}_\gamma^\beta = \mathcal{F}^{(1-\gamma)\beta}$, the Kolmogorov-Nagumo energy can be expressed as
\begin{eqnarray}
\mathcal{U}^\beta_\gamma &=& \mathcal{F}^\beta_\gamma + \frac{1}{\beta} \mathcal{R}_\gamma^\beta\nonumber\\
&=& \mathcal{F}^{\beta'} + \beta \, \frac{\mathcal{F}^{\beta'} - \mathcal{F}^\beta}{\beta'-\beta} \nonumber\\
&=& \frac{\beta' \mathcal{F}^{\beta'} - \beta \mathcal{F}^\beta}
{\beta'-\beta}\, .
\end{eqnarray}
By denoting free entropy as $\Psi^\beta = - \beta \mathcal{F}^\beta = \mathcal{S}^\beta - \beta \mathcal{U}^\beta$ we find that
\begin{equation}
\mathcal{U}_\gamma^\beta = - \frac{\Psi^{\beta'}-\Psi^\beta}{\beta'-\beta}~.
\end{equation}
Again, in the limit $\gamma \rightarrow 0$ we recover the classic relationship
\begin{equation}
\mathcal{U}^\beta = - \left(\frac{\partial \Psi^\beta}{\partial \beta}\right)\, .
\end{equation}

\section{Connection with multifractal thermodynamics}
\label{sec:multifractals}

\rev{
Let us now focus on the connection of Kolmogorov-Nagumo averages with multifractals. The multifractal analysis provides a powerful tool for investigating the self-similarity of complex systems, including physical and chemical systems \cite{mandelbrot1989multifractal}, weather forecast \cite{lovejoy2018weather} or financial systems \cite{calvet2008multifractal}. It has been shown that the R\'{e}nyi entropy plays a crucial role in the theory of multifractals \cite{jizba2004}.
We will show that the presented framework based on Kolmogorov-Nagumo averages establishes a connection between two distinct multifractal formalisms.}

\subsection*{Fundamentals in multifractal theory}

\rev{%We start with a brief overview of the standard multifractal formalism. 
Following the standard approach in the theory of multifractals \cite{harte}, let us consider a physical system whose state space is parcelled into distinct regions $k_i(s)$ indexed by $i\in\mathcal{I}$, with the partition depending on a characteristic scale $s$. Such partitions can be studied with respect to positional, spatiotemporal, or energetic state space. Consider the probability of observing the system within region $k_i(s)$, which is denoted by $p_i(s)$. Let us assume that this probability observes a scaling property of the form
\begin{equation}\label{eq:scaling}
p_i(s) = \frac{s^{\alpha_i}}{z(s)}\,,
\end{equation} 
where $\alpha_i$ is a scaling exponent and $z(s) = \sum_i s^{\alpha_i}$ is a normalization constant.
%Here we introduce the notation $p_i(s) \propto s^{\alpha_i}$ meaning that
%\begin{equation}
%\lim_{s \rightarrow 0}
%\frac{\log(p_i(s))}{\log %s} = \alpha_i + c \,,
%\end{equation}
%where we assume that the %imit
%\begin{equation}
%c = \lim_{s \rightarrow %0} \frac{\log z(s)}{\log s}
%\end{equation}
%exists and is a finite number that can be eventually incorporated into $\alpha_i$. Thus, without loss of generality, we focus in the case that $c=0$, or equivalently, $\sum_i s^{\alpha_i}=1$.
}

\rev{
For small scales, i.e. $s \rightarrow 0$, let us assume that the frequency of the scaling exponent $\alpha_i$ is given by a continuous probability distribution, whose density $\rho$ has the form
\begin{equation}
\label{eq:multifrac}
\rho(\alpha,s) \mathrm{d} \alpha = c(\alpha) s^{-f(\alpha)} \mathrm{d} \alpha\, .
\end{equation}
Above, $c(\alpha)$ is a slowly varying function of $\alpha$ and $f(\alpha)$ is the so-called \emph{multifractal spectrum} of the system, being the fractal dimension of subset which scales
with exponent $\alpha$.This means that the number of sets $k_i(s)$ that have the scaling exponent $\alpha_j$ scale as 
\begin{equation}
\mathrm{card}\{k_i(s)|p_i(s) = s^{\alpha}/z(s)\}  = N(s) s^{-f(\alpha)}  
\end{equation}
where $\mathrm{card}$ denotes cardinality (i.e., the number of sets) and $N(s)$ is the normalization constant.}

\rev{
Systems that satisfy eq.~\ref{eq:multifrac} are called  multifractals, and the scaling exponent of the R\'{e}nyi entropy, denoted by $D_\gamma$, is known as the generalized dimension  \cite{harte}, i.e., \cite{halsey1986fractal}
\begin{equation}
 \lim_{s \rightarrow 0} \frac{(1-\gamma) R_\gamma(s)}{\ln s} = D_\gamma    \,.
\end{equation}
}

\subsection{Multifractals and R\'{e}nyi's entropy}

\rev{
A direct calculation shows that the R\'{e}nyi entropy can be rewritten as
\begin{equation}
R_\gamma(s) = \frac{1}{\gamma(1-\gamma)} \ln \frac{\int s^{(1-\gamma)\alpha} c(\alpha) s^{-f(\alpha)}  \mathrm{d} \alpha}{\left(\int s^{\alpha} c(\alpha) s^{-f(\alpha)}  \mathrm{d} \alpha\right)^{1-\gamma}} \,.
\end{equation}
In the limit of $s \rightarrow 0$, both integrals can be approximated by the steepest descent approximation, i.e., it is possible to find a value $\alpha_\gamma$ (resp. $\alpha_1$) that maximizes the exponents in the integrals. To this end, we define
\begin{eqnarray}
\alpha_\gamma &=& \arg\min_\alpha\big\{(1-\gamma) \alpha - f(\alpha)\big\}\,,\\
\alpha_1 &=& \arg\min_\alpha\big\{\alpha - f(\alpha)\big\}\,,
\end{eqnarray}
and obtain that, for small $s$, the R\'{e}nyi entropy can be approximated as
\begin{equation}
R_\gamma(s) =
\left[\frac{\alpha_\gamma-\alpha_1}{\gamma} - \frac{f(\alpha_\gamma)-(1-\gamma)f(\alpha_1)}{\gamma(1-\gamma)}  \right] \ln s + \pazocal{O}(1),
%\frac{1}{\gamma(1-\gamma)} \ln [c(\alpha_\gamma) s^{-f(\alpha_\gamma) +(1-\gamma)\alpha_\gamma}]
\end{equation}
where we omitted the constant term coming from the normalization function $c(\alpha)$. By introducing the L\'{e}gendre transform of $f(\alpha)$, 
\begin{equation}
\tau_\gamma = (1-\gamma)\alpha_\gamma - f(\alpha_\gamma)\, ,
\end{equation}
one can express the generalized dimension as 
\begin{equation}
R_\gamma = \frac{\tau_\gamma - (1-\gamma) \tau_1}{\gamma(1-\gamma)} \ln s + \pazocal{O}(1) \, .
\end{equation}
As a result, we find that the connection between the multifractal spectrum and generalized dimension can be established as
\begin{equation}
D_\gamma = \frac{\tau_\gamma - (1-\gamma) \tau_1}{\gamma} \, .
\end{equation}
Thus, by calculating R\'{e}nyi entropy, one can obtain the multifractal spectrum and vice versa. }

\subsection{Turbulence cascades}

\rev{
Let us now focus on an alternative approach to multifractal based on turbulence cascades on the scaling of the energy field~\cite{schertzer1997multifractal}. The fundamental assumption at the base of this approach is that the average energy $\epsilon(s)$ scales as
$\left\langle \epsilon(s)^{\gamma} \right\rangle \sim s^{M_\gamma(\epsilon)}
$, which can be formally written as
\begin{equation}
\lim_{s \rightarrow 0} \frac{\ln \left\langle \epsilon(s)^{\gamma} \right\rangle}{\ln s} = M_\gamma(\epsilon) = \gamma \langle \epsilon \rangle_\gamma = \gamma U_{\gamma} \,.
\end{equation}
}

\rev{
Interestingly, this approach can be connected with the previous one by noting that the distribution obtained from maximization of R\'enyi entropy with the constraint on the cumulant generating function leads to 
$p(\epsilon_i) \equiv \pi_i$, i.e., Boltzmann distribution \eqref{eq:equi}. By comparing Eqs. \eqref{eq:equi} and \eqref{eq:scaling}, we find the following correspondence:
\begin{equation}
\alpha_i \ln s = - \beta \epsilon_i \,.
\end{equation}
Therefore, $\epsilon_i = \alpha_i$ relates the energy to the characteristic scaling exponent, and $\beta = - \ln s$ connects the inverse temperature with the characteristic scale. Furthermore, the parameter $\gamma$ plays the role of the rescaling 
of the characteristic scale, where $\beta' = (1-\gamma) \beta$ leads to
\begin{equation}
s' = s^{1-\gamma} \,.
\end{equation}
This implies that using the $\gamma$-exponential Kolmogorov averages implies a change of the characteristic scale $s \mapsto s^{1-\gamma}$. Finally, we obtained that 
\begin{eqnarray}
\beta U_\beta^{\gamma}(s) &=& \frac{\tau_\gamma - \tau_1}{\gamma}\ln s + \pazocal{O}(1) \,,\\
\beta F_\beta^{\gamma}(s) &=& \frac{\tau_\gamma}{\gamma-1} \ln s + \pazocal{O}(1) \,,\\
M_\gamma = \gamma U_\gamma^\beta(s) &=& \tau_1 - \tau_\gamma + o(1) \,.
\end{eqnarray}
We note that the relation between cumulant generating function has been described in Ref.~\cite{schertzer1997multifractal}.}

\rev{
These results show that the connection between the two formalisms naturally leads to thermodynamics with exponential Kolmogorov-Nagumo averages. 
Our results imply that, for the case of multifractal systems, the maximization of R\'{e}nyi entropy under the constraint of Kolmogorov-Nagumo average naturally gives us the Boltzmann distribution, where the energy can be translated to the scaling exponent of the equilibrium distribution. 
}

%*\section{Non-equilibrium thermodynamics}
\section{Non-equilibrium thermodynamics}
\label{sec:non-equi}

In this section \rev{we investigate the thermodynamics of }
non-equilibrium systems subject to constraints in the form of Kolmogorov-Nagumo averages. 
We focus on their L\'{e}gendre structure, $H$-theorem, entropy production, and detailed and integrated fluctuation theorems.

\subsection*{L\'{e}gendre structure}
The L\'{e}gendre transform establishes several key relations in thermodynamics, connecting the internal energy, entropy, temperature, and Helmholtz free energy --- which is defined as $F = U - T S$. It also establishes a natural link between extensive variables (e.g., energy, entropy) with intensive variables (e.g., temperature). Interestingly, 
while the L\'{e}gendre transform arises naturally in the classic Boltzmann-Gibbs framework, it has been shown that a more general L\'{e}gendre structure can still be applied to more general scenarios~\cite{plastino1997} in the context of thermodynamics and~\cite{e25040678} in information geometry.  

Following this line of reasoning, we now derive thermodynamic relationships that lead directly to a generalized L\'{e}gendre structure. The L\'{e}gendre structure for the case of Kolmogorov-Nagumo averages has also been investigated \rev{by Scarfone, Matsuzoe, and Wada} in \cite{scarfone2022}. Let us start by calculating the change of Helmholtz free energy from the equilibrium distribution to an arbitrary state. Using Eq.~\eqref{eq:f2}, this can be expressed as
\begin{equation}
F_\gamma(p)-F_\gamma(\pi) = \Delta U_\gamma^\beta - \tfrac{1}{\beta} \Delta R_\gamma~.
\end{equation}
The difference between non-equilibrium and equilibrium internal energy can be expressed as 
\begin{eqnarray}
 U^\beta_\gamma(p) - U^\beta_\gamma(\pi) =
 \frac{1}{\beta \gamma} \ln  \frac{\sum_i p_i e^{\gamma \beta_0 \epsilon_i}}{\sum_i \pi_i e^{\gamma \beta \epsilon_i}} \nonumber\\
 = \frac{1}{\beta}{\gamma} \ln \left(1 + \frac{\sum_i (p_i-\pi_i) \pi_i^{-\gamma}}{\sum_i \pi_i^{1-\gamma}} \right)\nonumber\\
 =  \frac{1}{\beta \gamma} \ln \left(1+ \gamma \,  \nabla R_\gamma \cdot (p-\pi)\right)~,
\end{eqnarray}
where $\nabla R_\gamma$ is the vector of partial derivatives of the R\'{e}nyi entropy expressed in the equilibrium state, which can be written as
\begin{eqnarray}\label{eq:par}
\frac{\partial R_\gamma}{\partial \pi_i} = \frac{1}{\gamma \sum_i (\pi_i)^{1-\gamma}} \pi_i^{-\gamma}~.
%=\frac{1}{\gamma}\frac{e^{\gamma \beta \epsilon_i} }{\sum_k \pi_k e^{\gamma \beta \epsilon_k}}.
\end{eqnarray}
Thus, the free energy difference can be expressed as
\begin{align}
\label{eq:f1}
 F_\gamma(p)-F_\gamma(\pi) =  
\frac{1}{\beta} R_\gamma(\pi) - \frac{1}{\beta} R_\gamma(p) \nonumber\\
+  \frac{1}{\beta \gamma} \ln \left(1+ \gamma \,  \nabla R_\gamma \cdot (p-\pi)\right)\equiv \tfrac{1}{\beta} \mathcal{D}_{\gamma}(p || \pi),
\end{align}
where $\mathcal{D}_{\gamma}(p || \pi)$ is the R\'{e}nyi-Bregmann divergence~\cite{wong2018logarithmic}. 
Hence, the divergence between non-equilibrium and equilibrium distributions can be re-written as
\begin{eqnarray}
\mathcal{D}_{\gamma}(p || \pi) &=& R_\gamma(\pi) - R_\gamma(p)  + C(\nabla R_\gamma,(p-\pi))\nonumber\\
&=&R_\gamma(\pi) - R_\gamma(p) + \frac{1}{\gamma} \ln \sum p_i \Pi_i^{(\gamma)},
\end{eqnarray}
where $\Pi_i^{(\gamma)} = \frac{\pi_i^{-\gamma}}{\sum_i \pi_i^{1-\gamma}}$, and $C(x,y) = \frac{1}{\gamma} \ln \left(1+ \gamma \, x \cdot y\right)$ is a generalized \textit{link function}, which becomes the ordinary dot product $x \cdot y$ for $\gamma \rightarrow 0$. The fact that $\mathcal{D}_{\gamma}(p || \pi)\geq 0$ \cite{wong2019logarithmic} implies that 
$\Delta F_\gamma$ is always positive, and therefore the free energy is minimized by the equilibrium distribution --- generalizing the classical result to the cases of non-linear averages. The implications of this result for thermodynamic scenarios are developed in the next section.

To conclude, let us note that while the standard free energy is obtained by a regular L\'{e}gendre transform on the macroscopic level (i.e., as $F = U - T S$), the more general free energy $F_\gamma$ is obtained as a generalized L\'{e}gendre transform on the mesoscopic (i.e., probability) level. This can be seen by comparing Eq. \eqref{eq:f2}, where free energy difference is obtained as $\Delta F_\gamma = \Delta U_\gamma^\beta - \beta^{-1} \Delta R_\gamma$, with Eq.~\eqref{eq:f1}, where the free energy difference is obtained as \mbox{$\Delta F = \beta^{-1} \left[ - \Delta R_\gamma + \ln_\gamma(\nabla R_\gamma \cdot \Delta p) \right].$}

\subsection*{Second law and H-theorem}
Let us now focus on the case of non-equilibrium stochastic thermodynamics of systems that follow constraints given by Kolmogorov-Nagumo averages. Stochastic thermodynamics \cite{seifert2012stochastic} recently became an important topic of non-equilibrium statistical physics. While there have been several attempts to derive stochastic thermodynamics associated with generalized entropies (see, e.g., Refs.~\cite{frank2004nonlinear,PhysRevE.101.022114,Korbel_2021}), there have been no studies focused on the stochastic thermodynamics of Kolmogorov-Nagumo averages. 

Before starting, let us consider a general formula for a time derivative of a $\gamma$-exponential average:
\begin{eqnarray}\label{eq:der_av}
\frac{\dd}{\dd t} \langle Y(t) \rangle_\gamma = \frac{\dd}{\dd t} \left(\frac{1}{\gamma} \ln\sum_i p_i(t) e^{\gamma y_i(t)}\right)  \nonumber\\
=\frac{1}{\gamma} \frac{\sum_i \dot{p}_i(t) e^{\gamma y_i(t)} + \sum_i p_i(t) \gamma \dot{y}_i(t) e^{\gamma y_i(t)}}{\sum_i p_i(t) e^{\gamma y_i(t)}}
\end{eqnarray}
With this expression at our disposal, let's start investigating non-equilibrium thermodynamics.

We begin by focusing on the second law of thermodynamics. For this, let us use Eq.~\eqref{eq:der_av} to re-state the first law of thermodynamics for the total energy measured as a Kolmogorov-Nagumo average, which reads as follows:
\begin{equation}\label{eq:first_law}
\frac{\mathrm{d} U^\beta_\gamma}{\mathrm{d} t} = \dot{Q}^\beta_\gamma + \dot{W}^\beta_\gamma,
\end{equation}
where the work and heat flow into the system of interest are given by
\begin{equation}
\dot{W}^\beta_\gamma = \frac{\sum_i \dot{\epsilon}_i p_i(t) e^{\gamma \beta \epsilon_i}}{\sum_i p_i(t) e^{\gamma \beta \epsilon_i}}
\quad
\text{and}
\quad
\dot{Q}^\beta_\gamma = \frac{1}{\beta \gamma} \frac{\sum_i \dot{p}_i(t) e^{\gamma \beta \epsilon_i}}{\sum_i p_i(t) e^{\gamma \beta \epsilon_i}}.
\end{equation}
Using these expressions, we can now derive the second law of thermodynamics for the case of R\'{e}nyi entropy 
%with Kolmogorov-Nagumo exponential mean for the case of 
and an arbitrary non-equilibrium process driven by linear Markov dynamics. To this end, let us consider a standard master equation given by
\begin{equation}\label{eq:master_eq}
\dot{p}_i(t) = \sum_j \Big(w_{ij}(t) p_j(t) - w_{ji}(t) p_i(t)\Big),  
\end{equation}
\rev{where $\dot{p}_i(t):= \frac{\dd p_i(t)}{\dd t}$ is the time derivative of $p_i(t)$}, $w_{ij}$ is the transition rate between states $i$ and $j$. \rev{Let us consider a control protocol $\lambda(t)$ which controls the energy spectrum $\epsilon_i(t) \equiv \epsilon_i(\lambda(t))$. Let us also introduce the Boltzmann distribution as $\pi_i(t) = \frac{1}{Z} \exp(-\beta(\epsilon_i(t))$.  We assume that the transition rates satisfy satisfies detailed balance, i.e.
\begin{equation}\label{eq:detailed_balance}
\frac{w_{ij}(t)}{w_{ji}(t)} = \frac{\pi_i(t)}{\pi_j(t)} = e^{\beta (\epsilon_j(t) - \epsilon_i(t))}.
\end{equation}
}
%
%\rev{In the rest of this section, we assume that the energy spectrum $\epsilon_i$ is time-independent, i.e., $\dot{\epsilon}_i = 0$. Therefore, the master equation with transition rates satisfying the detailed balance describes the pure relaxation to equilibrium. }

The time derivative of entropy and heat flow can then be expressed as
\begin{align}
\dot{R}_\gamma &= \frac{1}{\gamma} \frac{\sum_i (p_i)^{-\gamma} \dot{p}_i}{\sum_i p_i^{1-\gamma}}\nonumber \\
&=\frac{1}{\gamma} \sum_{ij} (w_{ij} p_j - w_{ji} p_i) P^{(\gamma)}_i,\\
\beta \dot{Q}_\gamma^\beta &= \frac{1}{\gamma}\frac{\sum_i \dot{p}_i e^{\gamma \beta \epsilon_i}}{\sum_i p_i e^{\gamma \beta \epsilon_i}} \nonumber \\
&= \frac{1}{\gamma} \sum_{ij} (w_{ij} p_j - w_{ji} p_i) \Phi_i^{(\gamma)},
\end{align}
where we are using \rev{$P^{(\gamma)}_i(t) = \frac{p_i^{-\gamma}(t)}{\sum_i p_i^{1-\gamma}(t)}$} and \rev{$\Phi^{(\gamma)}_i(t) = \frac{\pi_i^{-\gamma}(t)}{\sum_i p_i(t) \pi_i^{-\gamma}(t)}$} as shorthand notations. \rev{In the rest of the text, we will often omit the explicit dependence on time for the sake of clarity.}
Then, the entropy production rate can be calculated as
\begin{eqnarray}\label{eq:epr}
\dot{\Sigma}_\gamma &:=& \dot{R}_\gamma - \beta \dot{Q}^\beta_\gamma = \frac{1}{\gamma} \sum_{ij} (w_{ij}p_j - w_{ji} p_i) (P_i^{(\gamma)} - \Phi_i^{(\gamma)})\nonumber\\
&=& \frac{1}{2\gamma} \sum_{ij} (w_{ij}p_j - w_{ji} p_i) ([P_i^{(\gamma)}-P_j^{(\gamma)}] - [\Phi_i^{(\gamma)}-\Phi_j^{(\gamma)}])\nonumber\\
\end{eqnarray}
Thus, it is possible to write the entropy production rate as
\begin{equation}\label{eq:ep}
\dot{\Sigma}_\gamma = \frac{1}{2} \sum_{ij} J_{ij} F_{ij}~,
\end{equation}
where $J_{ij} := w_{ij}p_j - w_{ji}p_i$ are the thermodynamic fluxes and $F_{ij} = \sigma^{bath}_{ij} + \sigma^{sys}_{ij}$ the thermodynamic forces with
\begin{eqnarray}
\sigma^{bath}_{ij} :&=& - \frac{1}{\gamma}\left(\Phi^{(\gamma)}_i - \Phi^{(\gamma)}_j\right) \nonumber\\
&=& \frac{\Phi_i^{(\gamma)}}{\gamma} \, \left[\exp\left(\gamma \ln \frac{w_{ij}}{w_{ji}}\right)-1\right], \\
\sigma^{sys}_{ij} :&=& \frac{1}{\gamma}\left(P^{(\gamma)}_i - P^{(\gamma)}_j\right)\nonumber\\
&=&  \frac{P_j^{(\gamma)}}{\gamma} \left[\exp\left(\gamma \ln \frac{p_j}{p_i}\right)-1\right].
\end{eqnarray}
For the case of $\gamma \rightarrow 0$, the thermodynamic force reduces to the ordinary thermodynamic force $F_{ij} = \ln \frac{w_{ij}p_j}{w_{ji}p_i}$.
Using this derivation, one can show that the entropy production satisfies the second law of thermodynamics, which in this case, reads as
\begin{equation}\label{eq:second_law}
\dot{\Sigma}_\gamma = \dot{R}_\gamma -  \beta \dot{Q}_\gamma^\beta
\geq 0.
\end{equation}

Let us \rev{now} define a generalized $H$-function as $H_\gamma(p):= \mathcal{D}_\gamma(p||\pi) = \beta (F_\gamma(p) - F_\gamma(\pi))$, where the second equality is based on Eq.~\eqref{eq:f1}. Then, one can show that \rev{for the case of pure relaxation, i.e., where $w_{ij}$ does not depend on time and satisfies the detailed balance 
\begin{equation}\label{eq:detailed_balance2}
\frac{w_{ij}}{w_{ji}} = \frac{\pi_i}{\pi_j} = e^{\beta (\epsilon_j - \epsilon_i)}.
\end{equation}
}
$H_\gamma$ fulfills a generalized H-theorem given by the fact that $H_\gamma(p)\geq 0$ and
\begin{equation}\label{eq:htheorem}
\dot{H}_\gamma = -\frac{1}{\gamma} \sum_{ij}(w_{ij}p_j - w_{ji}p_i) \left(P_i^{(\gamma)} - \Pi_i^{(\gamma)}\right) \leq 0~,
\end{equation}
with $\dot{H}_\gamma=0$ if and only if $p=\pi$. This implies that $H_\gamma(p)$ is a Lyapunov function of the dynamics, i.e., a non-negative quantity that monotonically decreases with time until equilibrium is attained. 
The proof of both the second law and H-theorem is provided in Appendix~\ref{app:Proof_2ndlaw}.

Interestingly, the H-function (i.e., the distance from equilibrium distribution) and the entropy production (i.e., the maximum amount of reversible work) generally differ if $\gamma\neq0$, contrary to the case of the standard Boltzmann-Gibbs framework ($\gamma=0$) where both are equal to the Kullback-Leibler divergence. 
The difference between $\dot{\Sigma}_\gamma$ (Eq. \eqref{eq:second_law}) and $\dot{H}_\gamma$ (Eq. \eqref{eq:htheorem}) is in the replacement of $\Phi_i^{(\gamma)}$ by $\Pi_i^{(\gamma)}$, i.e., a different averaging in the denominator. 
In fact, by combining the first and second laws of thermodynamics, one can find that
\begin{eqnarray}
\dot{W}^{\beta}_\gamma &=& \dot{F}_{\gamma}^{\beta} + \frac{1}{\beta}  \dot{\Sigma}_{\gamma}^{\beta}  = \frac{1}{\beta} (\dot{H}_\gamma^\beta+\dot{\Sigma}_\gamma^\beta)\nonumber\\
&=& \frac{1}{\beta \gamma} \sum_{ij} (w_{ij}p_j - w_{ji}p_i) \left(\Pi_i^{(\gamma)} - \Phi_i^{(\gamma)} \right) \nonumber\\
&=&\frac{1}{\beta \gamma} \sum_{ij} (w_{ij}p_j - w_{ji}p_i) \left(\frac{\pi_i^{-\gamma}}{\sum_i \pi_i \pi_i^{-\gamma}} - \frac{\pi_i^{-\gamma}}{\sum_i p_i \pi_i^{-\gamma}} \right)\nonumber\\
\end{eqnarray}

To conclude, it is important to note that both Eqs.~\eqref{eq:second_law} and \eqref{eq:htheorem} reveal a family of inequalities --- indexed by $\gamma$ --- that hold for any stochastic system whose dynamics are governed by a master equation (as in Eq.~\eqref{eq:master_eq}) and satisfy \rev{detailed} balance (as in Eq.~\eqref{eq:detailed_balance}). 
Said differently, our results reveal that any process following a master equation and satisfying \rev{detailed} balance will satisfy those inequalities for all values of $\gamma$. 
It is then the constraints of the system that determine \textit{which values of $\gamma$ are physically meaningful}: systems obeying linear constraints yield to a Shannon-type second law (with $\gamma=0$), while non-linear Kolmogorov-Nagumo constraints lead to a R\'{e}nyi-type second law (with $\gamma\neq0$).

\subsection*{Fluctuation theorems}
We now show that trajectory thermodynamics remains in the same functional form as in ordinary stochastic thermodynamics. 
\rev{
Throughout this section, we will be denoting trajectories and their functionals by bold symbols. We will be denoting probabilities of observing a trajectory by uppercase, bold symbol $\pmb{P}$, which is the probability of observing a state $x$ at time $t$ by a lowercase symbol $p$.
%, to avoid confusion.
}

\rev{
Let us consider a trajectory $\pmb{x} =(x_0,t_0;x_1,t_1,\dots;x_f,t_f)$ starting at $t_0$ and finishing at $t_f$.
We denote the trajectory state at time $t$ as $\pmb{x}(t)$. The trajectory dwells in state $\pmb{x}(t) = x_{i-1}$ for $t_{i-1} \leq t <t_{i}$ and then jumps to a state $x_{i}$ at time $t_i$.  Let us define $j(\pmb{x})$ as the number of trajectory jumps.
The probability of observing a trajectory $\pmb{x}$ can be obtained from the master equation \eqref{eq:master_eq} as
\begin{equation}
\pmb{P}[\pmb{x}] = p_{x_0}(t_0) \prod_{j \in j(\pmb{x}) } e^{-\int_{t_{j-1}}^{t_j}w_{x_{j-1} x_{j-1}}(\tau) \dd \tau} w_{x_j x_{j-1}}(t_j) \,.
\end{equation}
The energy corresponding to the state $x$ is controlled by a time-dependent protocol $\lambda(t)$, so $\epsilon_x(t) \equiv \epsilon_x(\lambda(t))$. Similarly to the previous section, we assume that the transition rates satisfy the detailed balance \eqref{eq:detailed_balance}.
}

\rev{
Let us now introduce an operation of time reversal $\tilde{t} = t_f-(t-t_0)$. The time-reversed trajectory can be defined as $\tilde{\pmb{x}}(t) = \pmb{x}(\tilde{t})= \pmb{x}(t_f-(t-t_0))$. This concept is crucial for understanding the connection between the irreversibility of mesoscopic systems and entropy production. By considering the time-reversed dynamics with time-reversed control protocol $\tilde{\lambda}(t)$, the probability of observing the time-reversed trajectory $\tilde{\pmb{x}}$ in the time-reversed dynamics determined by the time-reversed protocol $\tilde{\lambda}$ can be expressed as
\begin{equation}
\tilde{\pmb{P}}[\tilde{\pmb{x}}] = p_{x_f}(t_f) \prod_{j \in j(\pmb{x}) } e^{-\int_{t_{j-1}}^{t_{j}}w_{x_{j-1} x_{j-1}}(\tau) \dd \tau} w_{x_{j-1} x_{j}}(t_j)\,,
\end{equation}
where $\tilde{P}$ denotes the fact that the probability is calculated in the time-reversed dynamics.
By calculating the log ratio of the trajectory probabilities, we get that the waiting probabilities $e^{-\int_{t_{j-1}}^{t_{j}}w_{x_{j-1} x_{j-1}}(\tau) \dd \tau}$ cancel out and we end with
\begin{equation}
\ln \frac{\pmb{P}\big[\pmb{x}\big]}{\tilde{\pmb{P}}\big[\tilde{\pmb{x}}\big]} = \ln p\big(x(t_0)\big) - \ln p\big(x(t_f)\big) + \ln \prod_{j \in j(\pmb{x})} \frac{w_{x_{j}x_{j-1}}(t_j)}{w_{x_{j}x_{j-1}}(t_j)}\,.
\end{equation}
}

\rev{
Let us now define the entropy of state $x$ at time $t$ as the Hartley information of  $p_{x}(t)$, i.e. %, for a single time; one can define 
$s_x(t) = - \log p_x(t)$. From this definition, it is possible to define a trajectory entropy 
$\pmb{s}[\pmb{x}](t) \equiv s_{\pmb{x}(t)}(t)= - \ln p_{\pmb{x}(t)}(t)$.  By using the condition of detailed balance, we can express the log-ratio as 
\begin{equation}
\ln \frac{\pmb{P}\big[\pmb{x}\big]}{\tilde{\pmb{P}}\big[\tilde{\pmb{x}}\big]} = s_{x_f}(t_f) - s_{x_0}(t_0) + \sum_{j \in j(\pmb{x})} \beta\left(\epsilon_{x_j}(t_j) - \epsilon_{x_{j-1}}(t_j)\right) \,.
\end{equation}
%which is justified by the fact that the Kolmogorov-Nagumo exponential average of $s(x(t))$ gives back the R\'{e}nyi entropy~(see Section~\ref{sec:Kol_nag}).
Above, the first difference is equal to the change of the trajectory entropy from the beginning of the trajectory to the end of the trajectory, and is denoted by $\Delta s[{\pmb{x}}]:= s_{x_f}(t_f) - s_{x_0}(t_0)$. Please note that while $\Delta s$ formally depends on the trajectory, it actually only depends on the starting and ending point (and hence we don't use the bold symbol for it). 
The second term is equal to the $\beta$ times the heat exchanged with the reservoir during the trajectory and is denoted by $\pmb{q}[\pmb{x}]:= \sum_{j \in j(\pmb{x})} \left(\epsilon_{x_j}(t_j) - \epsilon_{x_{j-1}}(t_j)\right)$. Thus, the log ratio of forward and reversed probabilities is equal to 
\begin{equation}\label{eq:ss2}
\ln \frac{\pmb{P}\big[\pmb{x}\big]}{\tilde{\pmb{P}}\big[\tilde{\pmb{x}}\big]} = \Delta s[\pmb{x}]  + \beta \pmb{q}[\pmb{x}] = \pmb{\sigma}[\pmb{x}]\,,
\end{equation}
which is the trajectory entropy production.
}

\rev{To show the relation between trajectory quantities and ensemble quantities, we calculate the time derivative of the entropy production. The time derivative of trajectory entropy can be expressed as
\begin{equation}
\frac{\partial \pmb{s}[\pmb{x}](t)}{\partial t} = - \frac{\dot{p}_{\pmb{x}(t)}}{p_{\pmb{x}(t)}} - \sum_{j \in j(\pmb{x})} \delta(t-t_j) \ln \frac{p_{x_j}(t)}{p_{x_{j-1}}(t)} \,,
\end{equation}
where the first term is due to the change in the probability distribution, and the second term is due to trajectory jumps. Similarly, the time derivative of trajectory heat is  
\begin{eqnarray}
\frac{\partial \pmb{q}[\pmb{x}](t)}{\partial t} &=& \sum_{j \in j(\pmb{x})} \delta(t-t_j) (\epsilon_{x_j}(t)-\epsilon_{x_{j-1}}(t)) \nonumber\\
&=& \frac{1}{\beta} \sum_{j \in j(\pmb{x})} \delta(t-t_j) \ln \frac{w_{x_{j}x_{j-1}}(t)}{w_{x_{j-1}x_{j}}(t)} \,.
\end{eqnarray}
In both cases, the time derivative depends only on $x^- = \pmb{x}(t^-)$ and $x^+ = \pmb{x}(t^+)$. By introducing 
\begin{eqnarray}    
\dot{s}_{x^-x^+} &=& \log \frac{p_{x^+}(t)}{p_{x^-}(t)} \,,\\
\dot{q}_{x^-x^+} &=& \frac{1}{\beta} \log \frac{w_{x^+x^-}(t)}{w_{x^-x^+}(t)} \,,
\end{eqnarray}
the ensemble entropy production rate (Eq.~\eqref{eq:ep}) can be expressed as
\begin{eqnarray}
\dot{\Sigma}_\gamma = \frac{1}{2} \sum_{x^-x^+} J_{x^-x^+} \left( \Phi_{x^-}^{(\gamma)} \frac{\exp(\gamma \beta \dot{q}_{x^-x^+})-1}{\gamma}\right.\nonumber\\
\left. +P_{x^+}^{(\gamma)}\frac{\exp(\gamma  \dot{s}_{x^-x^+})-1}{\gamma} \right) \,.
\end{eqnarray}
In this case, the relation between trajectory quantities and ensemble quantities is not so straightforward due to more complicated averaging.}

\rev{Let us now focus on another aspect of the entropy production, i.e., the measure of irreversibility}.
 \rev{We define a Kolmogorov-Nagumo average of all trajectories for a functional $\pmb{G}\big[\pmb{x}\big]$ defined for each probability as}
\begin{equation}
\big\langle \langle \pmb{G} \rangle \big\rangle_\gamma = \ln_\gamma \int \mathcal{D} \pmb{x} \pmb{P}\big[\pmb{x}\big] \exp_\gamma\!\big(\pmb{G}[\pmb{x}]\big)\,,
\end{equation}
where $\mathcal{D} \pmb{x}$ is the the path integral measure. Using this, one can find the ensemble entropy production as given by
\begin{eqnarray}
  \langle \langle \pmb{\sigma} \rangle \rangle_\gamma &=& \frac{1}{\gamma} \ln \int \mathcal{D} \pmb{x} \pmb{P}[\pmb{x}] \exp\left(\gamma  \ln \frac{\pmb{P}\big[\pmb{x}\big]}{\tilde{\pmb{P}}\big[\tilde{\pmb{x}}\big]}\right)\nonumber\\
  &=& \frac{1}{\gamma} \ln \int \mathcal{D} \pmb{x} (\pmb{P}[\pmb{x}])^{\gamma+1} (\tilde{\pmb{P}}[\tilde{\pmb{x}}])^{-\gamma}\nonumber\\
  &=& \mathfrak{D}_\gamma\big(\pmb{P}[{\pmb{x}}]\,||\,\tilde{\pmb{P}}[\tilde{\pmb{x}}]\big) \geq 0~,
\end{eqnarray}
where $\mathfrak{D}_\gamma(p||q) = \frac{1}{\gamma}\ln \int \dd x\,  p(x)^{\gamma+1}q(x)^{-\gamma}$ is the R\'{e}nyi-Csisz\'{a}r divergence.

Furthermore, one can use Eq.~\eqref{eq:ss2} to show the validity of a detailed fluctuation theorem that holds in the common form \cite{esposito2010three}. Let us define the probability of observing trajectory entropy production as 
\begin{equation}
P(\sigma) = \int \mathcal{D}\pmb{x} \pmb{P}[\pmb{x}] \delta\left(\sigma - \sigma_{\pmb{x}}\right)
\end{equation}
i.e., we sum over all trajectories that result in entropy production equal to $\sigma$. By a simple manipulation, we obtain
\begin{eqnarray}
P(\sigma) &=& \int \mathcal{D}\pmb{x} \pmb{P}[\pmb{x} ]\delta\left(\sigma - \ln \frac{\pmb{P}[\pmb{x}]}{\tilde{\pmb{P}}[\tilde{\pmb{x}}]}\right)\nonumber\\
&=&e^{\sigma} \int \mathcal{D}\pmb{x} \tilde{\pmb{P}}[\tilde{\pmb{x}}] \delta\left(-\Delta \sigma - \ln \frac{\tilde{\pmb{P}}[\tilde{\pmb{x}}]}{\pmb{P}[\pmb{x}]}\right)\nonumber\\
&=&e^{\Delta \sigma} \tilde{P}(-\Delta \sigma).
\end{eqnarray}
Here $\tilde{P}$ again denotes that probability is calculated  for the time-reversed dynamics. As a result, we obtain the detailed fluctuation theorem
\begin{equation}
\frac{P(\sigma)}{\tilde{P}(- \sigma)} = e^{\sigma}  ,
\end{equation}

Therefore, on the trajectory level, the relations remain exactly the same as in the case of the ordinary Shannon-Boltzmann-Gibbs framework.
Finally, the integrated fluctuation theorem can then be formulated as
\begin{equation}
\int \dd \sigma P(\sigma)  e^{\sigma}  = 1  \,,  
\end{equation}
where the integral takes place over the values of trajectory entropy production. 
An application of Jensen's inequality yields $\int \dd \sigma P(\sigma)  \sigma \geq 0$.

\section{Thermodynamic length}
\label{sec:thermo_length}

Thermodynamic length is a well-known metric that characterizes the distance between thermodynamic states. 
More specifically, this metric is related to the dissipation in a thermodynamic system due to finite-time transformations~\cite{salamon1983thermodynamic,nulton1985quasistatic}, and has important connections with the Jensen-Shannon divergence~\cite{bennett1976efficient}, Fisher information and Rao’s entropy differential metric~\cite{crooks2007measuring}. Therefore, thermodynamic length is of great interest for out-of-equilibrium analyses.

%Conventional observables in stochastic thermodynamics can be realized as elements in the space of probability distributions via line element. Within the information geometric framework,
%\cite{crooks2007measuring}
Let us consider a collection of thermodynamic states \rev{that can be parametrized by}  $(\theta_1,\dots,\theta_n)\in \Theta$. Then, the thermodynamic length of a path $\bm s(t)=\big(s_1(t),\dots,s_n(t)\big):[0,\tau]\to \Theta$ can be calculated as
\begin{equation}
\label{ther_length}
\mathcal{L} := \int_{0}^\tau \sqrt{\rev{\left(\frac{\dd \bm s}{\dd t}\right)^2}} \dd t = \int_{0}^\tau \sqrt{\sum_{ij} g_{ij} \dot{s}_i(t) \dot{s}_j(t)} \dd t ,
\end{equation}
%is known as statistical length, or \textit{thermodynamic length}. The $q_{i}$ correspond to probability distributions and 
\rev{$\dot{s}_i(t) := \frac{\dd s_i(t)}{\dd t}$ is the time derivative of $s_i(t)$ } where $g_{ij}$ \rev{is} a metric tensor corresponding to the well-known Fisher metric~\cite{amari2016information}, which is given by
\begin{equation}
\label{Fishermetric}
g_{ij}(q) = -\frac{\partial^2 \mathcal{D}(p||q)}{\partial p_i \partial q_j}\Bigg|_{p=q}.
\end{equation}
%
%Information geometry provides important tools to reason about both equilibrium and non-equilibrium thermodynamics \cite{PhysRevLett.99.100602,Kim2021,ito2018stochastic}, as well as critical phenomena \cite{janke2004information}. Information geometry has also recently found its applications in generalized thermodynamics \cite{naudts2008generalised,GHIKAS2018384,tempesta2019,korbel2019information,korbel2020information}, particularly in connection with R\'{e}nyi entropy \cite{de2016geometry,scarfone2020study}.
%
Above, $\mathcal{D}(p||q)$ corresponds to a \textit{divergence}, a central quantity in information geometry from which all other geometrical properties --- including the metric tensor~\eqref{Fishermetric} and connections --- can be derived. 
It is worth noting that there are two fundamental types of divergences: 
\begin{itemize}
    \item[(1)] Csisz\'{a}r divergences, of the form:
    $$\mathfrak{D}_{f,g}(p||q) = g\left(\sum_i q_i f\left(\frac{p_i}{q_i}\right)\right).$$
    \item[(2)] Bregmann divergence, of the form: 
    $$\mathcal{D}_{f,g}(p||q) = S_{f,g}(p) - S_{f,g}(q) - g\left(\langle \nabla S_{f,q}(q), p-q \rangle\right).$$
\end{itemize}
Both families of divergences are closely related to a trace-class generalized entropy $S_{f,g}(p) = g\left(\sum_i f(p_i)\right)$ \cite{hanel2012generalized}, \rev{where $f$ is an increasing function and $g$ is a concave function}. 
For the case of $g(x)=x$ and $f(x) = x \ln x$, both divergences reduce to the well-known Kullback-Leibler divergence. 
This shows that the Fisher-Rao metric corresponding to the Czisz\'{a}r divergence is equivalent to the Fisher metric corresponding to the Kullback-Leibler divergence~\cite{korbel2019information}. Furthermore, as elaborated in Eq.~\eqref{eq:f1} (see also Ref.~\cite{naudts2008generalised}), the Bregmann divergence corresponds to the difference of free energies --- i.e., the amount of reversible work between two states. These facts motivate us to focus on the R\'{e}nyi-Bregman divergence.

To recapitulate, Eq.~\eqref{eq:f1} shows that the R\'{e}nyi-Bregmann divergence can be expressed as
\begin{equation}
\mathcal{D}_\gamma(p||q)=  - R_\gamma(p) - R_\gamma(p,q),
\end{equation}
where $R_\gamma(p,q)$ is the R\'{e}nyi-Bregmann cross entropy given by
\begin{equation}
 R_\gamma(p,q) = -\frac{1}{\gamma} \ln \sum_i p_i q_i^{-\gamma} - \gamma R_\gamma(q) .
\end{equation}
This generalizes the standard relationship between Kullback-Leibler divergence, Shannon entropy, and cross-entropy $\mathcal{D}_0(p||q) = -H(p) - H_\text{cross}(p,q)$, which corresponds to the case of \rev{$\gamma=0$}.
%(which is found by applying L'H\^{o}spital's rule for $\gamma \rightarrow 0$, which leads to $R_0(p,q) = H_\text{cross}(p,q) = - \sum_i p_i \log q_i$). 
%, thus recovering the well-known relation . 
%Before further proceeding, let us mention several properties of $Q_i^{-\gamma}$. First, $Q_i^0 = 1$. Second, expected value w.r.t. $q_i$ can be expressed as
%\begin{equation}
%\langle Q^{-\gamma} \rangle_q = \sum_i q_i Q_i^{-\gamma} = \frac{\sum_i q_i q_i^{-\gamma}}{\sum_i q_i^{1-\gamma}} = 1    
%\end{equation}
%The derivative w.r.t. $q_j$ can be expressed as
%\begin{equation}
%\frac{\partial Q^{-\gamma}_i}{\partial q_j} = Q_i^{-\gamma} \left[(\gamma-1) Q_j^{-\gamma} - \gamma \frac{1}{q_j} \delta_{ij}\right].
%\end{equation}
%%%% to appendix
By leveraging the structure of the R\'{e}nyi-Bregmann divergence, one can find the following expression for the metric tensor valid for arbitrary $\gamma$:
\begin{equation}
g^{\gamma}_{ij}(q) = Q_i^{(\gamma)}Q_j^{(\gamma)} + \frac{Q_i^{(\gamma)}}{q_i} \delta_{ij} 
\end{equation}
where $Q^{(\gamma)}_i = \frac{q_i^{-\gamma}}{\sum_i q_i^{1-\gamma}}$ has been adopted for brevity.
Using this expression, one can then evaluate \rev{$\dd s^2$} as follows:
\begin{equation}
\label{eq:thdl}
\dd s^2 
%\sum_{ij} g_{ij}(q) \dot{q}_i \dot{q}_j = 
= \sum_i \frac{Q_i^{(\gamma)}}{q_i} (\dd q_i)^2 + \sum_{ij} Q_i^{(\gamma)}Q_j^{(\gamma)} \dd {q}_i \dd{q}_j.
\end{equation}
%In Appendix~\ref{app:ther_length}, we show that this expression can be rewritten as
%\begin{equation}\label{eq:unwieldy}
%\rev{\left(\frac{\dd s}{\dd t}\right)^2}=\frac{1}{\gamma-1} \sum_i \left[ (2\gamma -1) \frac{Q_i^{(\gamma)}}{q_i} (\dot{q}_i)^2 + \dot{Q}_i^{(\gamma)} \dot{q}_i \right]  .  
%\end{equation}
%The expression above is rather unwieldy but is well defined on the Shannonian limit, i.e. $\gamma \to 0$, hence we resort to a power expansion to extract the small $\gamma$ dependence of thermodynamic limit

%The expression in Eq.~\eqref{eq:unwieldy} %is rather unwieldy, however, 
%provides a generalization of the link between information geometry and thermodynamics. %Indeed, in~\cite{ito2018stochastic} via stochastic thermodynamic's master equation recasts the length element in its Shannonian limit, i.e. $\gamma \to 0$, in terms of the bath's entropy production rate and thermodynamic force. The integrated thermodynamics length presents a geometric constraint for the entropy change rate transfer as 
%In effect, with the aid of the master equation of stochastic thermodynamics, one may recast the \rev{expression for the thermodynamic length} in its Shannonian limit (i.e., $\gamma \to 0$) in terms of the bath's entropy production rate and thermodynamic force~\cite{ito2018stochastic}. In this limit, it can be seen that the thermodynamic length presents a geometric constraint for the entropy change rate transfer as a probability distribution transition to another.

\section{Conclusions}
\label{sec:conclusion}

This paper presents a description of the thermodynamics of systems that follow non-linear constraints in the form of Kolmogorov-Nagumo averages. 
%In practice, this means that the relevant constraints that characterise the system are non-linear. 
\rev{Our results provide a first step towards a deeper understanding of the thermodynamics of such systems, opening the door to the analysis of systems with long-range correlations and/or multifractal properties, which are naturally described by the R\'{e}nyi entropy \cite{jizba2019maximum,jizba2004}. 
Furthermore, recent applications of non-equilibrium thermodynamics based on R\'{e}nyi entropy (and consequently exponential Kolmogorov-Nagumo averages) to utility theory have been investigated \cite{ducuara2023maxwells}, opening the possibility to apply the presented formalism in the context of game theory.}

The thermodynamics of Kolmogorov-Nagumo averages was found to be naturally centered around the notion of R\'{e}nyi's entropy and a generalized L\'{e}gendre transform, which lead to a novel form of free entropy. Our results show that this free energy, in turn, is directly related to the entropy production of the system. The presented framework allows to extend of most thermodynamic relations to these non-linear systems --- including the second law of thermodynamics and fluctuation theorems --- if their relationships are adequately recast in terms of the deformed L\'{e}gendre transform, as summarized in Table \ref{tab:my_label}.

In the context of current attempts to apply generalized entropies known mainly from information theory in thermodynamics, it is worth mentioning that generalized entropies --- such as the R\'{e}nyi entropy --- do not only emerge in the case of systems under non-arithmetic means as constraints but also emerge in the description of systems that obey a non-linear master equation with ordinary arithmetic average %(e.g. as the constraint for the internal energy)
~\cite{Korbel_2021}. In the context of equilibrium thermodynamics, it has been argued --- perhaps \rev{surprisingly} --- that different entropies and constraints can lead to the same equilibrium distribution~\cite{Korbel21a}. Our results can be seen as extending these ideas for non-equilibrium processes, as they imply that different combinations of particular dynamics (i.e. the precise form of the Fokker-Planck/master equation), detailed balance (identifying stationary distribution with the equilibrium distribution), and energetic constraints can also lead to the same entropic functional. The precise choice of thermodynamic and dynamic relations depends on the particular choice of a physical system that one intends to describe. The investigation of how to better characterize classes of dynamics and constraints that lead to similar phenomena is an important topic that deserves further investigation in future work.

\begin{acknowledgments}
P.A.M. acknowledges support by JSPS KAKENHI Grant Number 23K168550001. J.K. acknowledges support by the Austrian Science Fund (FWF) project No. P 34994. F.R. was supported by the Fellowship Programme of the Institute of Cultural and Creative Industries of the University of Kent.
\end{acknowledgments}

\appendix

\section{Proof of the second law of thermodynamics and H-theorem for systems described by Kolmogorov-Nagumo averages}
\label{app:Proof_2ndlaw}

The time derivative of the entropy and heat rate over temperature can be found to be given by
\begin{align}
\dot{R}_\gamma &= \frac{1}{\gamma} \frac{\sum_i (p_i)^{-\gamma} \dot{p}_i}{\sum_i p_i^{1-\gamma}}\nonumber, \\
&=\frac{1}{\gamma} \sum_{ij} (w_{ij} p_j - w_{ji} p_i) \frac{p_i^{-\gamma}}{\sum_k p_k^{1-\gamma}} \\
\beta \dot{Q}_\gamma^\beta &= \frac{1}{\gamma}\frac{\sum_i \dot{p}_i e^{\gamma \beta \epsilon_i}}{\sum_i p_i e^{\gamma \beta \epsilon_i}} \nonumber \\
&= \frac{1}{\gamma} \sum_{ij} (w_{ij} p_j - w_{ji} p_i) \frac{e^{\gamma \beta \epsilon_i}}{\sum_k p_k e^{\gamma \beta \epsilon_k}}.
\end{align}
Above, the explicit dependence of $p_i(t)$  on time is omitted for simplicity. 
The entropy production rate can then be expressed as
\begin{eqnarray}
\dot{\Sigma}_\gamma &=& \frac{1}{\gamma} \sum_{ij}\left(w_{ij} p_j - w_{ji} p_i\right) \nonumber\\ && \times \left[ \frac{p_i^{-\gamma}}{\sum_k p_k^{1-\gamma}}\right. 
\left. -  \frac{e^{\gamma \beta \epsilon_i}}{\sum_k p_k e^{\gamma \beta \epsilon_k}} \right]\nonumber\\
&=& \frac{1}{\gamma} \sum_{ij} w_{ji}p_i \left(\frac{\pi_i}{\pi_{j}} \frac{p_j}{p_i} - 1\right)
\nonumber\\ && \times
\left[\frac{p_i^{-\gamma}}{\sum_k p_k^{1-\gamma}} - \frac{\sum_k \pi_k e^{\gamma \beta \epsilon_k}}{\sum_k p_k e^{\gamma \beta \epsilon_k}} 
 \frac{\pi_{i}^{-\gamma}}{\sum_{k} \pi_k^{1-\gamma}} \right].
\end{eqnarray}
Furthermore, by using Eq. \eqref{eq:par} one can find that 
\begin{equation}
  \frac{\sum_k \pi_k e^{\gamma \beta \epsilon_k}}{\sum_k p_k e^{\gamma \beta \epsilon_k}} 
 \frac{\pi_{i}^{-\gamma}}{\sum_{k} \pi_k^{1-\gamma}} = \frac{\pi_i^{-\gamma}}{\sum_k p_k \pi_k^{-\gamma} }~.
\end{equation}
By denoting $P^{(\gamma)}_i(t) = \frac{p_i^{-\gamma}(t)}{\sum_i p_i(t) p_i^{-\gamma}(t)}$ and $\Phi^{(\gamma)}_i(t) = \frac{\pi_i^{-\gamma}(t)}{\sum_i p_i(t) \pi_i^{-\gamma}(t)}$, we can then find that
\begin{equation}
\dot{\Sigma}_\gamma =\frac{1}{\gamma} \sum_{ij} w_{ji} p_i \left( \left[\frac{\Phi^{(\gamma)}_i}{\Phi^{(\gamma)}_j} \frac{P^{(\gamma)}_j}{P^{(\gamma)}_i}\right]^{-1/\gamma}-1\right)\left(P^{(\gamma)}_i - \Phi^{(\gamma)}_i\right)~.
\end{equation}
Similarly, an analogous derivation shows that the time derivative of H-function $\dot{H}_\gamma$ can be expressed as
\begin{equation}
\dot{H}_\gamma =- \frac{1}{\gamma} \sum_{ij} w_{ji} p_i \left( \left[\frac{\Pi^{(\gamma)}_i}{\Pi^{(\gamma)}_j} \frac{P^{(\gamma)}_j}{P^{(\gamma)}_i}\right]^{-1/\gamma}-1\right)\left(P^{(\gamma)}_i - \Pi^{(\gamma)}_i\right).
\end{equation}

To conclude the derivation, let us consider now, instead, consider a more general expression
\begin{equation}
\dot{Y}_\gamma =\frac{1}{\gamma} \sum_{ij} w_{ji} p_i \left( \left[\frac{\Upsilon^{(\gamma)}_i}{\Upsilon^{(\gamma)}_j} \frac{P^{(\gamma)}_j}{P^{(\gamma)}_i}\right]^{-1/\gamma}-1\right)\left(P^{(\gamma)}_i - \Upsilon^{(\gamma)}_i\right)~,
\end{equation}
and show that it is non-negative for $\Upsilon_i^{(\gamma)} = \Pi_i^{(\gamma)}$ and $\Upsilon_i^{(\gamma)} = \Phi_i^{(\gamma)}(t)$.
By using the inequality $\gamma (x^{-1/\gamma}-1) \geq \log(1/x)$ for $\gamma >0$, one can find that
\begin{equation}
\dot{Y}_\gamma \geq \frac{1}{\gamma^2} \sum_{ij} w_{ji} p_i \log\left(\frac{\Upsilon^{(\gamma)}_j}{\Upsilon^{(\gamma)}_i}\frac{P^{(\gamma)}_i}{P^{(\gamma)}_j}\right) (P^{(\gamma)}_i - \Upsilon^{(\gamma)}_i) =: \mathcal{Y}\, .
\end{equation}
For both 
$\Upsilon_i^{(\gamma)} = \Pi_i^{(\gamma)}$ or $\Upsilon_i^{(\gamma)} = \Phi_i^{(\gamma)}$ cases one can \rev{use the fact that the processes satisfies the detailed balance, i.e.,
\begin{eqnarray}
    \log\left(\frac{\Phi^{(\gamma)}_i}{\Phi^{(\gamma)}_j} \frac{P^{(\gamma)}_j}{P^{(\gamma)}_i}\right) &=& \frac{1}{\gamma} \log \left(\frac{\pi_j(t)}{\pi_i(t)}\frac{p_i(t)}{p_j(t)}\right)\nonumber\\ &=&  \frac{1}{\gamma} \log \left(\frac{w_{ji}(t)}{w_{ij}(t)}\frac{p_i(t)}{p_j(t)}\right) 
\end{eqnarray}
and 
\begin{eqnarray}
    \log\left(\frac{\Pi^{(\gamma)}_i}{\Pi^{(\gamma)}_j} \frac{P^{(\gamma)}_j}{P^{(\gamma)}_i}\right) &=& \frac{1}{\gamma} \log \left(\frac{\pi_j}{\pi_i}\frac{p_i(t)}{p_j(t)}\right)\nonumber\\ &=&  \frac{1}{\gamma} \log \left(\frac{w_{ji}}{w_{ij}}\frac{p_i(t)}{p_j(t)}\right) 
\end{eqnarray}
respectively}, and express $\mathcal{Y}$ as
\begin{equation}
\mathcal{Y} = \frac{1}{\gamma}\sum_{ij} 
 w_{ji} p_i \log\left(\frac{w_{ji} p_i}{w_{ij} p_j}\right) (P^{(\gamma)}_i - \Upsilon^{(\gamma)}_i)~.
\end{equation}
Then, by using $\log(1/x) \geq 1-x$, one can obtain the following inequality:
\begin{eqnarray}
\mathcal{Y} &\geq \frac{1}{\gamma}\sum_{ij} w_{ji} p_i (P^{(\gamma)}_i - \Upsilon^{(\gamma)}_i) - \frac{1}{\gamma}\sum_{ij} w_{ij} p_j (P^{(\gamma)}_i - \Upsilon^{(\gamma)}_i)\nonumber\\
&= -\frac{1}{\gamma}\sum_{ij} (w_{ij}p_j - w_{ji}p_i) (P_i^{(\gamma)}-\Upsilon_i^{(\gamma)}) ~,
\end{eqnarray}
which is equal to $-\dot{Y}_\gamma$. Thus, one finally finds that $\dot{Y}_\gamma \geq - \dot{Y}_\gamma$, which in turn leads to $ \dot{Y}_\gamma \geq 0$. This proves both the second law $(\dot{\Sigma}_\gamma \geq 0)$ and the H-theorem $(\dot{H}_\gamma \leq 0)$.

\bibliography{references}

\begin{thebibliography}{71}%
\makeatletter
\providecommand \@ifxundefined [1]{%
 \@ifx{#1\undefined}
}%
\providecommand \@ifnum [1]{%
 \ifnum #1\expandafter \@firstoftwo
 \else \expandafter \@secondoftwo
 \fi
}%
\providecommand \@ifx [1]{%
 \ifx #1\expandafter \@firstoftwo
 \else \expandafter \@secondoftwo
 \fi
}%
\providecommand \natexlab [1]{#1}%
\providecommand \enquote  [1]{``#1''}%
\providecommand \bibnamefont  [1]{#1}%
\providecommand \bibfnamefont [1]{#1}%
\providecommand \citenamefont [1]{#1}%
\providecommand \href@noop [0]{\@secondoftwo}%
\providecommand \href [0]{\begingroup \@sanitize@url \@href}%
\providecommand \@href[1]{\@@startlink{#1}\@@href}%
\providecommand \@@href[1]{\endgroup#1\@@endlink}%
\providecommand \@sanitize@url [0]{\catcode `\\12\catcode `\$12\catcode
  `\&12\catcode `\#12\catcode `\^12\catcode `\_12\catcode `\%12\relax}%
\providecommand \@@startlink[1]{}%
\providecommand \@@endlink[0]{}%
\providecommand \url  [0]{\begingroup\@sanitize@url \@url }%
\providecommand \@url [1]{\endgroup\@href {#1}{\urlprefix }}%
\providecommand \urlprefix  [0]{URL }%
\providecommand \Eprint [0]{\href }%
\providecommand \doibase [0]{http://dx.doi.org/}%
\providecommand \selectlanguage [0]{\@gobble}%
\providecommand \bibinfo  [0]{\@secondoftwo}%
\providecommand \bibfield  [0]{\@secondoftwo}%
\providecommand \translation [1]{[#1]}%
\providecommand \BibitemOpen [0]{}%
\providecommand \bibitemStop [0]{}%
\providecommand \bibitemNoStop [0]{.\EOS\space}%
\providecommand \EOS [0]{\spacefactor3000\relax}%
\providecommand \BibitemShut  [1]{\csname bibitem#1\endcsname}%
\let\auto@bib@innerbib\@empty
%</preamble>
\bibitem [{\citenamefont {A.N.}(1930)}]{Kolmogorov1930388}%
  \BibitemOpen
  \bibfield  {author} {\bibinfo {author} {\bibfnamefont {Kolmogorov}\
  \bibnamefont {A.N.}},\ }\bibfield  {title} {\enquote {\bibinfo {title} {Sur
  la notion de la moyenne},}\ }\href
  {https://www.scopus.com/inward/record.uri?eid=2-s2.0-0000971462&partnerID=40&md5=637f9c4cdf31d23c300c10c7a01625aa}
  {\bibfield  {journal} {\bibinfo  {journal} {Accad. Naz. Lincei Mem. Cl. Sci.
  Fis. Mat. Natur. Sez.}\ }\textbf {\bibinfo {volume} {12}},\ \bibinfo {pages}
  {388 – 391} (\bibinfo {year} {1930})}\BibitemShut {NoStop}%
\bibitem [{\citenamefont {M.}(1930)}]{Nagumo193071}%
  \BibitemOpen
  \bibfield  {author} {\bibinfo {author} {\bibfnamefont {Nagumo}\ \bibnamefont
  {M.}},\ }\bibfield  {title} {\enquote {\bibinfo {title} {\"{U}ber eine klasse
  der mittelwerte},}\ }\href
  {https://www.scopus.com/inward/record.uri?eid=2-s2.0-0002048382&partnerID=40&md5=8d7ed44e465c965171ebc05708c23a22}
  {\bibfield  {journal} {\bibinfo  {journal} {Japanese Journal of Mathematics}\
  }\textbf {\bibinfo {volume} {7}},\ \bibinfo {pages} {71 – 79} (\bibinfo
  {year} {1930})}\BibitemShut {NoStop}%
\bibitem [{\citenamefont {De~Finetti}(1931)}]{de1931sul}%
  \BibitemOpen
  \bibfield  {author} {\bibinfo {author} {\bibfnamefont {Bruno}\ \bibnamefont
  {De~Finetti}},\ }\href@noop {} {\emph {\bibinfo {title} {Sul concetto di
  media}}}\ (\bibinfo  {publisher} {Istituto italiano degli attuari},\ \bibinfo
  {year} {1931})\BibitemShut {NoStop}%
\bibitem [{\citenamefont {Jessen}(1931)}]{jessen}%
  \BibitemOpen
  \bibfield  {author} {\bibinfo {author} {\bibfnamefont {B.}~\bibnamefont
  {Jessen}},\ }\bibfield  {title} {\enquote {\bibinfo {title} {\"{U}ber die
  verallgemeinerung des arithmetischen mittels},}\ }\href@noop {} {\bibfield
  {journal} {\bibinfo  {journal} {Acta Sci. Math.}\ }\textbf {\bibinfo {volume}
  {4}},\ \bibinfo {pages} {108--116} (\bibinfo {year} {1931})}\BibitemShut
  {NoStop}%
\bibitem [{\citenamefont {KITAGAWA}(1934)}]{kitagawa1934some}%
  \BibitemOpen
  \bibfield  {author} {\bibinfo {author} {\bibfnamefont {Tosio}\ \bibnamefont
  {KITAGAWA}},\ }\bibfield  {title} {\enquote {\bibinfo {title} {On some class
  of weighted means},}\ }\href@noop {} {\bibfield  {journal} {\bibinfo
  {journal} {Proceedings of the Physico-Mathematical Society of Japan. 3rd
  Series}\ }\textbf {\bibinfo {volume} {16}},\ \bibinfo {pages} {117--126}
  (\bibinfo {year} {1934})}\BibitemShut {NoStop}%
\bibitem [{\citenamefont {Acz{\'e}l}(1948)}]{aczel1948mean}%
  \BibitemOpen
  \bibfield  {author} {\bibinfo {author} {\bibfnamefont {J{\'a}nos}\
  \bibnamefont {Acz{\'e}l}},\ }\bibfield  {title} {\enquote {\bibinfo {title}
  {On mean values},}\ }\href@noop {} {\bibfield  {journal} {\bibinfo  {journal}
  {Bulletin of the American Mathematical Society}\ }\textbf {\bibinfo {volume}
  {54}},\ \bibinfo {pages} {392--400} (\bibinfo {year} {1948})}\BibitemShut
  {NoStop}%
\bibitem [{\citenamefont {Fodor}\ and\ \citenamefont
  {Roubens}(1995)}]{fodor1995meaningfulness}%
  \BibitemOpen
  \bibfield  {author} {\bibinfo {author} {\bibfnamefont {J{\'a}nos}\
  \bibnamefont {Fodor}}\ and\ \bibinfo {author} {\bibfnamefont {Marc}\
  \bibnamefont {Roubens}},\ }\bibfield  {title} {\enquote {\bibinfo {title} {On
  meaningfulness of means},}\ }\href@noop {} {\bibfield  {journal} {\bibinfo
  {journal} {Journal of Computational and Applied Mathematics}\ }\textbf
  {\bibinfo {volume} {64}},\ \bibinfo {pages} {103--115} (\bibinfo {year}
  {1995})}\BibitemShut {NoStop}%
\bibitem [{\citenamefont {Nielsen}\ and\ \citenamefont
  {Sun}(2016)}]{Nielsen16}%
  \BibitemOpen
  \bibfield  {author} {\bibinfo {author} {\bibfnamefont {Frank}\ \bibnamefont
  {Nielsen}}\ and\ \bibinfo {author} {\bibfnamefont {Ke}~\bibnamefont {Sun}},\
  }\bibfield  {title} {\enquote {\bibinfo {title} {Guaranteed bounds on
  information-theoretic measures of univariate mixtures using piecewise
  log-sum-exp inequalities},}\ }\href@noop {} {\bibfield  {journal} {\bibinfo
  {journal} {Entropy}\ }\textbf {\bibinfo {volume} {18}} (\bibinfo {year}
  {2016})}\BibitemShut {NoStop}%
\bibitem [{\citenamefont {Porcu}\ \emph {et~al.}(2009)\citenamefont {Porcu},
  \citenamefont {Mateu},\ and\ \citenamefont {Christakos}}]{Porcu2009}%
  \BibitemOpen
  \bibfield  {author} {\bibinfo {author} {\bibfnamefont {Emilio}\ \bibnamefont
  {Porcu}}, \bibinfo {author} {\bibfnamefont {Jorge}\ \bibnamefont {Mateu}}, \
  and\ \bibinfo {author} {\bibfnamefont {George}\ \bibnamefont {Christakos}},\
  }\bibfield  {title} {\enquote {\bibinfo {title} {Quasi-arithmetic means of
  covariance functions with potential applications to space–time data},}\
  }\href@noop {} {\bibfield  {journal} {\bibinfo  {journal} {Journal of
  Multivariate Analysis}\ }\textbf {\bibinfo {volume} {100}},\ \bibinfo {pages}
  {1830--1844} (\bibinfo {year} {2009})}\BibitemShut {NoStop}%
\bibitem [{\citenamefont {Klement}\ \emph {et~al.}(1999)\citenamefont
  {Klement}, \citenamefont {Mesiar},\ and\ \citenamefont {Pap}}]{Klement99}%
  \BibitemOpen
  \bibfield  {author} {\bibinfo {author} {\bibfnamefont {Erich~Peter}\
  \bibnamefont {Klement}}, \bibinfo {author} {\bibfnamefont {Radko}\
  \bibnamefont {Mesiar}}, \ and\ \bibinfo {author} {\bibfnamefont {Endre}\
  \bibnamefont {Pap}},\ }\bibfield  {title} {\enquote {\bibinfo {title} {Quasi-
  and pseudo-inverses of monotone functions, and the construction of
  t-norms},}\ }\href@noop {} {\bibfield  {journal} {\bibinfo  {journal} {Fuzzy
  Sets and Systems}\ }\textbf {\bibinfo {volume} {104}},\ \bibinfo {pages}
  {3--13} (\bibinfo {year} {1999})},\ \bibinfo {note} {triangular
  Norms}\BibitemShut {NoStop}%
\bibitem [{\citenamefont {R{\'e}nyi}(1976)}]{renyi1976}%
  \BibitemOpen
  \bibfield  {author} {\bibinfo {author} {\bibfnamefont {A.}~\bibnamefont
  {R{\'e}nyi}},\ }\href@noop {} {\emph {\bibinfo {title} {Selected Papers of
  Alfr{\'e}d R{\'e}nyi}}},\ \bibinfo {series} {Selected Papers of Alfr{\'e}d
  R{\'e}nyi}\ No.~\bibinfo {number} {2}\ (\bibinfo  {publisher} {Akad{\'e}miai
  Kiad{\'o}},\ \bibinfo {year} {1976})\BibitemShut {NoStop}%
\bibitem [{\citenamefont {St{\'e}phan}\ \emph {et~al.}(2014)\citenamefont
  {St{\'e}phan}, \citenamefont {Inglis}, \citenamefont {Fendley},\ and\
  \citenamefont {Melko}}]{stephan2014geometric}%
  \BibitemOpen
  \bibfield  {author} {\bibinfo {author} {\bibfnamefont {Jean-Marie}\
  \bibnamefont {St{\'e}phan}}, \bibinfo {author} {\bibfnamefont {Stephen}\
  \bibnamefont {Inglis}}, \bibinfo {author} {\bibfnamefont {Paul}\ \bibnamefont
  {Fendley}}, \ and\ \bibinfo {author} {\bibfnamefont {Roger~G}\ \bibnamefont
  {Melko}},\ }\bibfield  {title} {\enquote {\bibinfo {title} {Geometric mutual
  information at classical critical points},}\ }\href@noop {} {\bibfield
  {journal} {\bibinfo  {journal} {Physical review letters}\ }\textbf {\bibinfo
  {volume} {112}},\ \bibinfo {pages} {127204} (\bibinfo {year}
  {2014})}\BibitemShut {NoStop}%
\bibitem [{\citenamefont {St\'ephan}(2014)}]{ShannonRenyiQuantumSpin}%
  \BibitemOpen
  \bibfield  {author} {\bibinfo {author} {\bibfnamefont {Jean-Marie}\
  \bibnamefont {St\'ephan}},\ }\bibfield  {title} {\enquote {\bibinfo {title}
  {{Shannon and R\'enyi mutual information in quantum critical spin chains}},}\
  }\href@noop {} {\bibfield  {journal} {\bibinfo  {journal} {Phys. Rev. B}\
  }\textbf {\bibinfo {volume} {90}},\ \bibinfo {pages} {045424} (\bibinfo
  {year} {2014})}\BibitemShut {NoStop}%
\bibitem [{\citenamefont {Jizba}\ \emph {et~al.}(2016)\citenamefont {Jizba},
  \citenamefont {Ma}, \citenamefont {Hayes},\ and\ \citenamefont
  {Dunningham}}]{jizba2016one}%
  \BibitemOpen
  \bibfield  {author} {\bibinfo {author} {\bibfnamefont {Petr}\ \bibnamefont
  {Jizba}}, \bibinfo {author} {\bibfnamefont {Yue}\ \bibnamefont {Ma}},
  \bibinfo {author} {\bibfnamefont {Anthony}\ \bibnamefont {Hayes}}, \ and\
  \bibinfo {author} {\bibfnamefont {Jacob~A}\ \bibnamefont {Dunningham}},\
  }\bibfield  {title} {\enquote {\bibinfo {title} {One-parameter class of
  uncertainty relations based on entropy power},}\ }\href@noop {} {\bibfield
  {journal} {\bibinfo  {journal} {Physical Review E}\ }\textbf {\bibinfo
  {volume} {93}},\ \bibinfo {pages} {060104} (\bibinfo {year}
  {2016})}\BibitemShut {NoStop}%
\bibitem [{\citenamefont {Dong}(2016)}]{Dong:2016fnf}%
  \BibitemOpen
  \bibfield  {author} {\bibinfo {author} {\bibfnamefont {Xi}~\bibnamefont
  {Dong}},\ }\bibfield  {title} {\enquote {\bibinfo {title} {{The Gravity Dual
  of Renyi Entropy}},}\ }\href@noop {} {\bibfield  {journal} {\bibinfo
  {journal} {Nature Commun.}\ }\textbf {\bibinfo {volume} {7}},\ \bibinfo
  {pages} {12472} (\bibinfo {year} {2016})}\BibitemShut {NoStop}%
\bibitem [{\citenamefont {Barrella}\ \emph {et~al.}(2013)\citenamefont
  {Barrella}, \citenamefont {Dong}, \citenamefont {Hartnoll},\ and\
  \citenamefont {Martin}}]{Barrella:2013wja}%
  \BibitemOpen
  \bibfield  {author} {\bibinfo {author} {\bibfnamefont {Taylor}\ \bibnamefont
  {Barrella}}, \bibinfo {author} {\bibfnamefont {Xi}~\bibnamefont {Dong}},
  \bibinfo {author} {\bibfnamefont {Sean~A.}\ \bibnamefont {Hartnoll}}, \ and\
  \bibinfo {author} {\bibfnamefont {Victoria~L.}\ \bibnamefont {Martin}},\
  }\bibfield  {title} {\enquote {\bibinfo {title} {{Holographic entanglement
  beyond classical gravity}},}\ }\href@noop {} {\bibfield  {journal} {\bibinfo
  {journal} {JHEP}\ }\textbf {\bibinfo {volume} {09}},\ \bibinfo {pages} {109}
  (\bibinfo {year} {2013})}\BibitemShut {NoStop}%
\bibitem [{\citenamefont {Jizba}\ and\ \citenamefont
  {Korbel}(2019)}]{jizba2019maximum}%
  \BibitemOpen
  \bibfield  {author} {\bibinfo {author} {\bibfnamefont {Petr}\ \bibnamefont
  {Jizba}}\ and\ \bibinfo {author} {\bibfnamefont {Jan}\ \bibnamefont
  {Korbel}},\ }\bibfield  {title} {\enquote {\bibinfo {title} {{Maximum entropy
  principle in statistical inference: Case for non-Shannonian entropies}},}\
  }\href@noop {} {\bibfield  {journal} {\bibinfo  {journal} {Physical review
  letters}\ }\textbf {\bibinfo {volume} {122}},\ \bibinfo {pages} {120601}
  (\bibinfo {year} {2019})}\BibitemShut {NoStop}%
\bibitem [{\citenamefont {St\'ephan}\ \emph {et~al.}(2014)\citenamefont
  {St\'ephan}, \citenamefont {Inglis}, \citenamefont {Fendley},\ and\
  \citenamefont {Melko}}]{GeometricMutInf}%
  \BibitemOpen
  \bibfield  {author} {\bibinfo {author} {\bibfnamefont {Jean-Marie}\
  \bibnamefont {St\'ephan}}, \bibinfo {author} {\bibfnamefont {Stephen}\
  \bibnamefont {Inglis}}, \bibinfo {author} {\bibfnamefont {Paul}\ \bibnamefont
  {Fendley}}, \ and\ \bibinfo {author} {\bibfnamefont {Roger~G.}\ \bibnamefont
  {Melko}},\ }\bibfield  {title} {\enquote {\bibinfo {title} {{Geometric Mutual
  Information at Classical Critical Points}},}\ }\href@noop {} {\bibfield
  {journal} {\bibinfo  {journal} {Phys. Rev. Lett.}\ }\textbf {\bibinfo
  {volume} {112}},\ \bibinfo {pages} {127204} (\bibinfo {year}
  {2014})}\BibitemShut {NoStop}%
\bibitem [{\citenamefont {Iaconis}\ \emph {et~al.}(2013)\citenamefont
  {Iaconis}, \citenamefont {Inglis}, \citenamefont {Kallin},\ and\
  \citenamefont {Melko}}]{DetectingPhaseTwithRenyi}%
  \BibitemOpen
  \bibfield  {author} {\bibinfo {author} {\bibfnamefont {Jason}\ \bibnamefont
  {Iaconis}}, \bibinfo {author} {\bibfnamefont {Stephen}\ \bibnamefont
  {Inglis}}, \bibinfo {author} {\bibfnamefont {Ann~B.}\ \bibnamefont {Kallin}},
  \ and\ \bibinfo {author} {\bibfnamefont {Roger~G.}\ \bibnamefont {Melko}},\
  }\bibfield  {title} {\enquote {\bibinfo {title} {Detecting classical phase
  transitions with renyi mutual information},}\ }\href@noop {} {\bibfield
  {journal} {\bibinfo  {journal} {Phys. Rev. B}\ }\textbf {\bibinfo {volume}
  {87}},\ \bibinfo {pages} {195134} (\bibinfo {year} {2013})}\BibitemShut
  {NoStop}%
\bibitem [{\citenamefont {Zaletel}\ \emph {et~al.}(2011)\citenamefont
  {Zaletel}, \citenamefont {Bardarson},\ and\ \citenamefont
  {Moore}}]{PhysRevLett.107.020402}%
  \BibitemOpen
  \bibfield  {author} {\bibinfo {author} {\bibfnamefont {Michael~P.}\
  \bibnamefont {Zaletel}}, \bibinfo {author} {\bibfnamefont {Jens~H.}\
  \bibnamefont {Bardarson}}, \ and\ \bibinfo {author} {\bibfnamefont {Joel~E.}\
  \bibnamefont {Moore}},\ }\bibfield  {title} {\enquote {\bibinfo {title}
  {{Logarithmic Terms in Entanglement Entropies of 2D Quantum Critical Points
  and Shannon Entropies of Spin Chains}},}\ }\href@noop {} {\bibfield
  {journal} {\bibinfo  {journal} {Phys. Rev. Lett.}\ }\textbf {\bibinfo
  {volume} {107}},\ \bibinfo {pages} {020402} (\bibinfo {year}
  {2011})}\BibitemShut {NoStop}%
\bibitem [{\citenamefont {Jizba}\ and\ \citenamefont
  {Arimitsu}(2004{\natexlab{a}})}]{jizba2004}%
  \BibitemOpen
  \bibfield  {author} {\bibinfo {author} {\bibfnamefont {Petr}\ \bibnamefont
  {Jizba}}\ and\ \bibinfo {author} {\bibfnamefont {Toshihico}\ \bibnamefont
  {Arimitsu}},\ }\bibfield  {title} {\enquote {\bibinfo {title} {The world
  according to rényi: thermodynamics of multifractal systems},}\ }\href
  {\doibase https://doi.org/10.1016/j.aop.2004.01.002} {\bibfield  {journal}
  {\bibinfo  {journal} {Annals of Physics}\ }\textbf {\bibinfo {volume}
  {312}},\ \bibinfo {pages} {17--59} (\bibinfo {year}
  {2004}{\natexlab{a}})}\BibitemShut {NoStop}%
\bibitem [{\citenamefont {Jizba}\ and\ \citenamefont
  {Arimitsu}(2004{\natexlab{b}})}]{jizba04b}%
  \BibitemOpen
  \bibfield  {author} {\bibinfo {author} {\bibfnamefont {P.}~\bibnamefont
  {Jizba}}\ and\ \bibinfo {author} {\bibfnamefont {T.}~\bibnamefont
  {Arimitsu}},\ }\bibfield  {title} {\enquote {\bibinfo {title} {Observability
  of r\'enyi's entropy},}\ }\href@noop {} {\bibfield  {journal} {\bibinfo
  {journal} {Physical Review E}\ }\textbf {\bibinfo {volume} {69}},\ \bibinfo
  {pages} {026128} (\bibinfo {year} {2004}{\natexlab{b}})}\BibitemShut
  {NoStop}%
\bibitem [{\citenamefont {Jizba}\ \emph {et~al.}(2012)\citenamefont {Jizba},
  \citenamefont {Kleinert},\ and\ \citenamefont {Shefaat}}]{jizba2012renyi}%
  \BibitemOpen
  \bibfield  {author} {\bibinfo {author} {\bibfnamefont {Petr}\ \bibnamefont
  {Jizba}}, \bibinfo {author} {\bibfnamefont {Hagen}\ \bibnamefont {Kleinert}},
  \ and\ \bibinfo {author} {\bibfnamefont {Mohammad}\ \bibnamefont {Shefaat}},\
  }\bibfield  {title} {\enquote {\bibinfo {title} {R{\'e}nyi’s information
  transfer between financial time series},}\ }\href@noop {} {\bibfield
  {journal} {\bibinfo  {journal} {Physica A: Statistical Mechanics and its
  Applications}\ }\textbf {\bibinfo {volume} {391}},\ \bibinfo {pages}
  {2971--2989} (\bibinfo {year} {2012})}\BibitemShut {NoStop}%
\bibitem [{\citenamefont {Jizba}\ and\ \citenamefont
  {Korbel}(2014)}]{jizba2014multifractal}%
  \BibitemOpen
  \bibfield  {author} {\bibinfo {author} {\bibfnamefont {Petr}\ \bibnamefont
  {Jizba}}\ and\ \bibinfo {author} {\bibfnamefont {Jan}\ \bibnamefont
  {Korbel}},\ }\bibfield  {title} {\enquote {\bibinfo {title} {Multifractal
  diffusion entropy analysis: Optimal bin width of probability histograms},}\
  }\href@noop {} {\bibfield  {journal} {\bibinfo  {journal} {Physica A:
  Statistical Mechanics and its Applications}\ }\textbf {\bibinfo {volume}
  {413}},\ \bibinfo {pages} {438--458} (\bibinfo {year} {2014})}\BibitemShut
  {NoStop}%
\bibitem [{\citenamefont {Thurner}\ \emph {et~al.}(2018)\citenamefont
  {Thurner}, \citenamefont {Hanel},\ and\ \citenamefont
  {Klimek}}]{thurner2018}%
  \BibitemOpen
  \bibfield  {author} {\bibinfo {author} {\bibfnamefont {Stefan}\ \bibnamefont
  {Thurner}}, \bibinfo {author} {\bibfnamefont {Rudolf}\ \bibnamefont {Hanel}},
  \ and\ \bibinfo {author} {\bibfnamefont {Peter}\ \bibnamefont {Klimek}},\
  }\href@noop {} {\emph {\bibinfo {title} {Introduction to the theory of
  complex systems}}}\ (\bibinfo  {publisher} {Oxford University Press},\
  \bibinfo {year} {2018})\BibitemShut {NoStop}%
\bibitem [{\citenamefont {Borges}(2004)}]{borges2004}%
  \BibitemOpen
  \bibfield  {author} {\bibinfo {author} {\bibfnamefont {Ernesto~P}\
  \bibnamefont {Borges}},\ }\bibfield  {title} {\enquote {\bibinfo {title} {A
  possible deformed algebra and calculus inspired in nonextensive
  thermostatistics},}\ }\href@noop {} {\bibfield  {journal} {\bibinfo
  {journal} {Physica A: Statistical Mechanics and its Applications}\ }\textbf
  {\bibinfo {volume} {340}},\ \bibinfo {pages} {95--101} (\bibinfo {year}
  {2004})}\BibitemShut {NoStop}%
\bibitem [{\citenamefont {Czachor}(2020)}]{Czachor2020}%
  \BibitemOpen
  \bibfield  {author} {\bibinfo {author} {\bibfnamefont {Marek}\ \bibnamefont
  {Czachor}},\ }\bibfield  {title} {\enquote {\bibinfo {title} {Unifying
  aspects of generalized calculus},}\ }\href {\doibase 10.3390/e22101180}
  {\bibfield  {journal} {\bibinfo  {journal} {Entropy}\ }\textbf {\bibinfo
  {volume} {22}} (\bibinfo {year} {2020}),\ 10.3390/e22101180}\BibitemShut
  {NoStop}%
\bibitem [{\citenamefont {Beck}\ and\ \citenamefont
  {Sch\"{o}gl}(1993)}]{beck1993}%
  \BibitemOpen
  \bibfield  {author} {\bibinfo {author} {\bibfnamefont {Christian}\
  \bibnamefont {Beck}}\ and\ \bibinfo {author} {\bibfnamefont {Friedrich}\
  \bibnamefont {Sch\"{o}gl}},\ }\href@noop {} {\emph {\bibinfo {title}
  {Thermodynamics of Chaotic Systems: An Introduction}}},\ Cambridge Nonlinear
  Science Series\ (\bibinfo  {publisher} {Cambridge University Press},\
  \bibinfo {year} {1993})\BibitemShut {NoStop}%
\bibitem [{\citenamefont {Abe}(2003)}]{Abe03}%
  \BibitemOpen
  \bibfield  {author} {\bibinfo {author} {\bibfnamefont {Sumiyoshi}\
  \bibnamefont {Abe}},\ }\bibfield  {title} {\enquote {\bibinfo {title}
  {Geometry of escort distributions},}\ }\href@noop {} {\bibfield  {journal}
  {\bibinfo  {journal} {Phys. Rev. E}\ }\textbf {\bibinfo {volume} {68}},\
  \bibinfo {pages} {031101} (\bibinfo {year} {2003})}\BibitemShut {NoStop}%
\bibitem [{\citenamefont {Bercher}(2011)}]{Bercher11}%
  \BibitemOpen
  \bibfield  {author} {\bibinfo {author} {\bibfnamefont {J.‐F.}\ \bibnamefont
  {Bercher}},\ }\bibfield  {title} {\enquote {\bibinfo {title} {On escort
  distributions, q‐gaussians and fisher information},}\ }\href@noop {}
  {\bibfield  {journal} {\bibinfo  {journal} {AIP Conference Proceedings}\
  }\textbf {\bibinfo {volume} {1305}},\ \bibinfo {pages} {208--215} (\bibinfo
  {year} {2011})}\BibitemShut {NoStop}%
\bibitem [{\citenamefont {Frank}(2004)}]{frank2004nonlinear}%
  \BibitemOpen
  \bibfield  {author} {\bibinfo {author} {\bibfnamefont {TD}~\bibnamefont
  {Frank}},\ }\bibfield  {title} {\enquote {\bibinfo {title} {On a nonlinear
  master equation and the haken-kelso-bunz model},}\ }\href@noop {} {\bibfield
  {journal} {\bibinfo  {journal} {Journal of Biological Physics}\ }\textbf
  {\bibinfo {volume} {30}},\ \bibinfo {pages} {139--159} (\bibinfo {year}
  {2004})}\BibitemShut {NoStop}%
\bibitem [{\citenamefont {Korbel}\ and\ \citenamefont
  {Wolpert}(2021)}]{Korbel_2021}%
  \BibitemOpen
  \bibfield  {author} {\bibinfo {author} {\bibfnamefont {Jan}\ \bibnamefont
  {Korbel}}\ and\ \bibinfo {author} {\bibfnamefont {David~H}\ \bibnamefont
  {Wolpert}},\ }\bibfield  {title} {\enquote {\bibinfo {title} {Stochastic
  thermodynamics and fluctuation theorems for non-linear systems},}\
  }\href@noop {} {\bibfield  {journal} {\bibinfo  {journal} {New Journal of
  Physics}\ }\textbf {\bibinfo {volume} {23}},\ \bibinfo {pages} {033049}
  (\bibinfo {year} {2021})}\BibitemShut {NoStop}%
\bibitem [{\citenamefont {Czachor}\ and\ \citenamefont
  {Naudts}(2002)}]{czachor}%
  \BibitemOpen
  \bibfield  {author} {\bibinfo {author} {\bibfnamefont {Marek}\ \bibnamefont
  {Czachor}}\ and\ \bibinfo {author} {\bibfnamefont {Jan}\ \bibnamefont
  {Naudts}},\ }\bibfield  {title} {\enquote {\bibinfo {title} {Thermostatistics
  based on kolmogorov–nagumo averages: unifying framework for extensive and
  nonextensive generalizations},}\ }\href@noop {} {\bibfield  {journal}
  {\bibinfo  {journal} {Physics Letters A}\ }\textbf {\bibinfo {volume}
  {298}},\ \bibinfo {pages} {369--374} (\bibinfo {year} {2002})}\BibitemShut
  {NoStop}%
\bibitem [{\citenamefont {Morales}\ and\ \citenamefont
  {Rosas}(2021)}]{PhysRevResearch.3.033216}%
  \BibitemOpen
  \bibfield  {author} {\bibinfo {author} {\bibfnamefont {Pablo~A.}\
  \bibnamefont {Morales}}\ and\ \bibinfo {author} {\bibfnamefont {Fernando~E.}\
  \bibnamefont {Rosas}},\ }\bibfield  {title} {\enquote {\bibinfo {title}
  {Generalization of the maximum entropy principle for curved statistical
  manifolds},}\ }\href@noop {} {\bibfield  {journal} {\bibinfo  {journal}
  {Phys. Rev. Research}\ }\textbf {\bibinfo {volume} {3}},\ \bibinfo {pages}
  {033216} (\bibinfo {year} {2021})}\BibitemShut {NoStop}%
\bibitem [{\citenamefont {Tsallis}(1988)}]{tsallis1988possible}%
  \BibitemOpen
  \bibfield  {author} {\bibinfo {author} {\bibfnamefont {Constantino}\
  \bibnamefont {Tsallis}},\ }\bibfield  {title} {\enquote {\bibinfo {title}
  {Possible generalization of boltzmann-gibbs statistics},}\ }\href@noop {}
  {\bibfield  {journal} {\bibinfo  {journal} {Journal of statistical physics}\
  }\textbf {\bibinfo {volume} {52}},\ \bibinfo {pages} {479--487} (\bibinfo
  {year} {1988})}\BibitemShut {NoStop}%
\bibitem [{\citenamefont {Bagci}\ and\ \citenamefont
  {Tirnakli}(2009)}]{BAGCI20093230}%
  \BibitemOpen
  \bibfield  {author} {\bibinfo {author} {\bibfnamefont {G.~Baris}\
  \bibnamefont {Bagci}}\ and\ \bibinfo {author} {\bibfnamefont {Ugur}\
  \bibnamefont {Tirnakli}},\ }\bibfield  {title} {\enquote {\bibinfo {title}
  {On the way towards a generalized entropy maximization procedure},}\
  }\href@noop {} {\bibfield  {journal} {\bibinfo  {journal} {Physics Letters
  A}\ }\textbf {\bibinfo {volume} {373}},\ \bibinfo {pages} {3230--3234}
  (\bibinfo {year} {2009})}\BibitemShut {NoStop}%
\bibitem [{\citenamefont {Baez}(2022)}]{baez}%
  \BibitemOpen
  \bibfield  {author} {\bibinfo {author} {\bibfnamefont {John~C.}\ \bibnamefont
  {Baez}},\ }\bibfield  {title} {\enquote {\bibinfo {title} {R\'{e}nyi entropy
  and free energy},}\ }\href@noop {} {\bibfield  {journal} {\bibinfo  {journal}
  {Entropy}\ }\textbf {\bibinfo {volume} {24}},\ \bibinfo {pages} {706}
  (\bibinfo {year} {2022})}\BibitemShut {NoStop}%
\bibitem [{\citenamefont {Plastino}\ and\ \citenamefont
  {Plastino}(1997)}]{plastino1997}%
  \BibitemOpen
  \bibfield  {author} {\bibinfo {author} {\bibfnamefont {A}~\bibnamefont
  {Plastino}}\ and\ \bibinfo {author} {\bibfnamefont {AR}~\bibnamefont
  {Plastino}},\ }\bibfield  {title} {\enquote {\bibinfo {title} {On the
  universality of thermodynamics' legendre transform structure},}\ }\href@noop
  {} {\bibfield  {journal} {\bibinfo  {journal} {Physics Letters A}\ }\textbf
  {\bibinfo {volume} {226}},\ \bibinfo {pages} {257--263} (\bibinfo {year}
  {1997})}\BibitemShut {NoStop}%
\bibitem [{\citenamefont {Scarfone}\ \emph {et~al.}(2016)\citenamefont
  {Scarfone}, \citenamefont {Matsuzoe},\ and\ \citenamefont
  {Wada}}]{scarfone2022}%
  \BibitemOpen
  \bibfield  {author} {\bibinfo {author} {\bibfnamefont {A.M.}\ \bibnamefont
  {Scarfone}}, \bibinfo {author} {\bibfnamefont {H.}~\bibnamefont {Matsuzoe}},
  \ and\ \bibinfo {author} {\bibfnamefont {T.}~\bibnamefont {Wada}},\
  }\bibfield  {title} {\enquote {\bibinfo {title} {Consistency of the structure
  of legendre transform in thermodynamics with the kolmogorov–nagumo
  average},}\ }\href {\doibase https://doi.org/10.1016/j.physleta.2016.07.012}
  {\bibfield  {journal} {\bibinfo  {journal} {Physics Letters A}\ }\textbf
  {\bibinfo {volume} {380}},\ \bibinfo {pages} {3022--3028} (\bibinfo {year}
  {2016})}\BibitemShut {NoStop}%
\bibitem [{\citenamefont {Wong}(2018)}]{wong2018logarithmic}%
  \BibitemOpen
  \bibfield  {author} {\bibinfo {author} {\bibfnamefont {Ting-Kam~Leonard}\
  \bibnamefont {Wong}},\ }\bibfield  {title} {\enquote {\bibinfo {title}
  {{Logarithmic divergences from optimal transport and R{\'e}nyi geometry}},}\
  }\href@noop {} {\bibfield  {journal} {\bibinfo  {journal} {Information
  Geometry}\ }\textbf {\bibinfo {volume} {1}},\ \bibinfo {pages} {39--78}
  (\bibinfo {year} {2018})}\BibitemShut {NoStop}%
\bibitem [{\citenamefont {Wong}\ and\ \citenamefont
  {Yang}(2019)}]{wong2019logarithmic}%
  \BibitemOpen
  \bibfield  {author} {\bibinfo {author} {\bibfnamefont {Ting-Kam~Leonard}\
  \bibnamefont {Wong}}\ and\ \bibinfo {author} {\bibfnamefont {Jiaowen}\
  \bibnamefont {Yang}},\ }\bibfield  {title} {\enquote {\bibinfo {title}
  {{Logarithmic divergences: geometry and interpretation of curvature}},}\ }in\
  \href@noop {} {\emph {\bibinfo {booktitle} {International Conference on
  Geometric Science of Information}}}\ (\bibinfo {organization} {Springer},\
  \bibinfo {year} {2019})\ pp.\ \bibinfo {pages} {413--422}\BibitemShut
  {NoStop}%
\bibitem [{\citenamefont {Peng}\ \emph {et~al.}(2020)\citenamefont {Peng},
  \citenamefont {Qian},\ and\ \citenamefont {Hong}}]{PhysRevE.101.022114}%
  \BibitemOpen
  \bibfield  {author} {\bibinfo {author} {\bibfnamefont {Liangrong}\
  \bibnamefont {Peng}}, \bibinfo {author} {\bibfnamefont {Hong}\ \bibnamefont
  {Qian}}, \ and\ \bibinfo {author} {\bibfnamefont {Liu}\ \bibnamefont
  {Hong}},\ }\bibfield  {title} {\enquote {\bibinfo {title} {Thermodynamics of
  markov processes with nonextensive entropy and free energy},}\ }\href
  {\doibase 10.1103/PhysRevE.101.022114} {\bibfield  {journal} {\bibinfo
  {journal} {Phys. Rev. E}\ }\textbf {\bibinfo {volume} {101}},\ \bibinfo
  {pages} {022114} (\bibinfo {year} {2020})}\BibitemShut {NoStop}%
\bibitem [{\citenamefont {Enciso}\ \emph {et~al.}(2019)\citenamefont {Enciso},
  \citenamefont {Gun}, \citenamefont {Ruiz},\ and\ \citenamefont
  {Razzitte}}]{Enciso_2019}%
  \BibitemOpen
  \bibfield  {author} {\bibinfo {author} {\bibfnamefont {Luciano}\ \bibnamefont
  {Enciso}}, \bibinfo {author} {\bibfnamefont {Marcelo}\ \bibnamefont {Gun}},
  \bibinfo {author} {\bibfnamefont {María~Sol}\ \bibnamefont {Ruiz}}, \ and\
  \bibinfo {author} {\bibfnamefont {Adrián~C}\ \bibnamefont {Razzitte}},\
  }\bibfield  {title} {\enquote {\bibinfo {title} {Entropy in multifractal non
  equilibrium structures of dielectric breakdown},}\ }\href@noop {} {\bibfield
  {journal} {\bibinfo  {journal} {Journal of Statistical Mechanics: Theory and
  Experiment}\ }\textbf {\bibinfo {volume} {2019}},\ \bibinfo {pages} {094011}
  (\bibinfo {year} {2019})}\BibitemShut {NoStop}%
\bibitem [{\citenamefont {Amari}(2016)}]{amari2016information}%
  \BibitemOpen
  \bibfield  {author} {\bibinfo {author} {\bibfnamefont {Shun-ichi}\
  \bibnamefont {Amari}},\ }\href@noop {} {\emph {\bibinfo {title} {{Information
  geometry and its applications}}}},\ Vol.\ \bibinfo {volume} {194}\ (\bibinfo
  {publisher} {Springer},\ \bibinfo {year} {2016})\BibitemShut {NoStop}%
\bibitem [{\citenamefont {Naudts}(2008)}]{naudts2008generalised}%
  \BibitemOpen
  \bibfield  {author} {\bibinfo {author} {\bibfnamefont {Jan}\ \bibnamefont
  {Naudts}},\ }\bibfield  {title} {\enquote {\bibinfo {title} {Generalised
  exponential families and associated entropy functions},}\ }\href@noop {}
  {\bibfield  {journal} {\bibinfo  {journal} {Entropy}\ }\textbf {\bibinfo
  {volume} {10}},\ \bibinfo {pages} {131--149} (\bibinfo {year}
  {2008})}\BibitemShut {NoStop}%
\bibitem [{\citenamefont {Korbel}\ \emph {et~al.}(2019)\citenamefont {Korbel},
  \citenamefont {Hanel},\ and\ \citenamefont
  {Thurner}}]{korbel2019information}%
  \BibitemOpen
  \bibfield  {author} {\bibinfo {author} {\bibfnamefont {Jan}\ \bibnamefont
  {Korbel}}, \bibinfo {author} {\bibfnamefont {Rudolf}\ \bibnamefont {Hanel}},
  \ and\ \bibinfo {author} {\bibfnamefont {Stefan}\ \bibnamefont {Thurner}},\
  }\bibfield  {title} {\enquote {\bibinfo {title} {Information geometric
  duality of $\phi$-deformed exponential families},}\ }\href@noop {} {\bibfield
   {journal} {\bibinfo  {journal} {Entropy}\ }\textbf {\bibinfo {volume}
  {21}},\ \bibinfo {pages} {112} (\bibinfo {year} {2019})}\BibitemShut
  {NoStop}%
\bibitem [{\citenamefont {Korbel}\ \emph {et~al.}(2020)\citenamefont {Korbel},
  \citenamefont {Hanel},\ and\ \citenamefont
  {Thurner}}]{korbel2020information}%
  \BibitemOpen
  \bibfield  {author} {\bibinfo {author} {\bibfnamefont {Jan}\ \bibnamefont
  {Korbel}}, \bibinfo {author} {\bibfnamefont {Rudolf}\ \bibnamefont {Hanel}},
  \ and\ \bibinfo {author} {\bibfnamefont {Stefan}\ \bibnamefont {Thurner}},\
  }\bibfield  {title} {\enquote {\bibinfo {title} {Information geometry of
  scaling expansions of non-exponentially growing configuration spaces},}\
  }\href@noop {} {\bibfield  {journal} {\bibinfo  {journal} {The European
  Physical Journal Special Topics}\ }\textbf {\bibinfo {volume} {229}},\
  \bibinfo {pages} {787--807} (\bibinfo {year} {2020})}\BibitemShut {NoStop}%
\bibitem [{\citenamefont {Eguchi}\ \emph {et~al.}(2016)\citenamefont {Eguchi},
  \citenamefont {Komori},\ and\ \citenamefont {Ohara}}]{eguchi2016information}%
  \BibitemOpen
  \bibfield  {author} {\bibinfo {author} {\bibfnamefont {Shinto}\ \bibnamefont
  {Eguchi}}, \bibinfo {author} {\bibfnamefont {Osamu}\ \bibnamefont {Komori}},
  \ and\ \bibinfo {author} {\bibfnamefont {Atsumi}\ \bibnamefont {Ohara}},\
  }\bibfield  {title} {\enquote {\bibinfo {title} {Information geometry
  associated with generalized means},}\ }in\ \href@noop {} {\emph {\bibinfo
  {booktitle} {Information Geometry and its Applications IV}}}\ (\bibinfo
  {organization} {Springer},\ \bibinfo {year} {2016})\ pp.\ \bibinfo {pages}
  {279--295}\BibitemShut {NoStop}%
\bibitem [{\citenamefont {Acz{\'e}l}\ \emph {et~al.}(1989)\citenamefont
  {Acz{\'e}l}, \citenamefont {Acz{\'e}l}, \citenamefont {Acz{\'e}l},\ and\
  \citenamefont {Dhombres}}]{aczel1989functional}%
  \BibitemOpen
  \bibfield  {author} {\bibinfo {author} {\bibfnamefont {J{\'a}nos}\
  \bibnamefont {Acz{\'e}l}}, \bibinfo {author} {\bibfnamefont
  {J{\'a}nos~Dezoo}\ \bibnamefont {Acz{\'e}l}}, \bibinfo {author}
  {\bibfnamefont {Joseph}\ \bibnamefont {Acz{\'e}l}}, \ and\ \bibinfo {author}
  {\bibfnamefont {J}~\bibnamefont {Dhombres}},\ }\href@noop {} {\emph {\bibinfo
  {title} {Functional equations in several variables}}},\ \bibinfo {number}
  {31}\ (\bibinfo  {publisher} {Cambridge university press},\ \bibinfo {year}
  {1989})\BibitemShut {NoStop}%
\bibitem [{\citenamefont {Hardy}\ \emph {et~al.}(1952)\citenamefont {Hardy},
  \citenamefont {Littlewood}, \citenamefont {P{\'o}lya}, \citenamefont
  {P{\'o}lya} \emph {et~al.}}]{hardy1952inequalities}%
  \BibitemOpen
  \bibfield  {author} {\bibinfo {author} {\bibfnamefont {Godfrey~Harold}\
  \bibnamefont {Hardy}}, \bibinfo {author} {\bibfnamefont {John~Edensor}\
  \bibnamefont {Littlewood}}, \bibinfo {author} {\bibfnamefont {George}\
  \bibnamefont {P{\'o}lya}}, \bibinfo {author} {\bibfnamefont {Gy{\"o}rgy}\
  \bibnamefont {P{\'o}lya}},  \emph {et~al.},\ }\href@noop {} {\emph {\bibinfo
  {title} {Inequalities}}}\ (\bibinfo  {publisher} {Cambridge university
  press},\ \bibinfo {year} {1952})\BibitemShut {NoStop}%
\bibitem [{\citenamefont {Touchette}(2009)}]{touchette2009large}%
  \BibitemOpen
  \bibfield  {author} {\bibinfo {author} {\bibfnamefont {Hugo}\ \bibnamefont
  {Touchette}},\ }\bibfield  {title} {\enquote {\bibinfo {title} {The large
  deviation approach to statistical mechanics},}\ }\href@noop {} {\bibfield
  {journal} {\bibinfo  {journal} {Physics Reports}\ }\textbf {\bibinfo {volume}
  {478}},\ \bibinfo {pages} {1--69} (\bibinfo {year} {2009})}\BibitemShut
  {NoStop}%
\bibitem [{\citenamefont {Hartley}(1928)}]{hartley1928}%
  \BibitemOpen
  \bibfield  {author} {\bibinfo {author} {\bibfnamefont {Ralph~VL}\
  \bibnamefont {Hartley}},\ }\bibfield  {title} {\enquote {\bibinfo {title}
  {Transmission of information 1},}\ }\href@noop {} {\bibfield  {journal}
  {\bibinfo  {journal} {Bell System technical journal}\ }\textbf {\bibinfo
  {volume} {7}},\ \bibinfo {pages} {535--563} (\bibinfo {year}
  {1928})}\BibitemShut {NoStop}%
\bibitem [{\citenamefont {Ince}(2017)}]{ince2017measuring}%
  \BibitemOpen
  \bibfield  {author} {\bibinfo {author} {\bibfnamefont {Robin~AA}\
  \bibnamefont {Ince}},\ }\bibfield  {title} {\enquote {\bibinfo {title}
  {Measuring multivariate redundant information with pointwise common change in
  surprisal},}\ }\href@noop {} {\bibfield  {journal} {\bibinfo  {journal}
  {Entropy}\ }\textbf {\bibinfo {volume} {19}},\ \bibinfo {pages} {318}
  (\bibinfo {year} {2017})}\BibitemShut {NoStop}%
\bibitem [{\citenamefont {Campbell}(1965)}]{campbell65}%
  \BibitemOpen
  \bibfield  {author} {\bibinfo {author} {\bibfnamefont {L.L.}\ \bibnamefont
  {Campbell}},\ }\bibfield  {title} {\enquote {\bibinfo {title} {A coding
  theorem and r\'{e}nyi's entropy},}\ }\href@noop {} {\bibfield  {journal}
  {\bibinfo  {journal} {Information and Control}\ }\textbf {\bibinfo {volume}
  {8}},\ \bibinfo {pages} {423 -- 429} (\bibinfo {year} {1965})}\BibitemShut
  {NoStop}%
\bibitem [{\citenamefont {Burg}(1972)}]{Burg}%
  \BibitemOpen
  \bibfield  {author} {\bibinfo {author} {\bibfnamefont {John~Parker}\
  \bibnamefont {Burg}},\ }\bibfield  {title} {\enquote {\bibinfo {title} {The
  relationship between maximum entropy spectra and maximum likelihood
  spectra},}\ }\href {\doibase 10.1190/1.1440265} {\bibfield  {journal}
  {\bibinfo  {journal} {Geophysics}\ }\textbf {\bibinfo {volume} {37}},\
  \bibinfo {pages} {375--376} (\bibinfo {year} {1972})}\BibitemShut {NoStop}%
\bibitem [{\citenamefont {Mandelbrot}(1989)}]{mandelbrot1989multifractal}%
  \BibitemOpen
  \bibfield  {author} {\bibinfo {author} {\bibfnamefont {Beno{\^\i}t~B}\
  \bibnamefont {Mandelbrot}},\ }\bibfield  {title} {\enquote {\bibinfo {title}
  {Multifractal measures, especially for the geophysicist},}\ }\href@noop {}
  {\bibfield  {journal} {\bibinfo  {journal} {Fractals in geophysics}\ ,\
  \bibinfo {pages} {5--42}} (\bibinfo {year} {1989})}\BibitemShut {NoStop}%
\bibitem [{\citenamefont {Lovejoy}\ and\ \citenamefont
  {Schertzer}(2018)}]{lovejoy2018weather}%
  \BibitemOpen
  \bibfield  {author} {\bibinfo {author} {\bibfnamefont {Shaun}\ \bibnamefont
  {Lovejoy}}\ and\ \bibinfo {author} {\bibfnamefont {Daniel}\ \bibnamefont
  {Schertzer}},\ }\href@noop {} {\emph {\bibinfo {title} {The weather and
  climate: emergent laws and multifractal cascades}}}\ (\bibinfo  {publisher}
  {Cambridge University Press},\ \bibinfo {year} {2018})\BibitemShut {NoStop}%
\bibitem [{\citenamefont {Calvet}\ and\ \citenamefont
  {Fisher}(2008)}]{calvet2008multifractal}%
  \BibitemOpen
  \bibfield  {author} {\bibinfo {author} {\bibfnamefont {Laurent~E}\
  \bibnamefont {Calvet}}\ and\ \bibinfo {author} {\bibfnamefont {Adlai~J}\
  \bibnamefont {Fisher}},\ }\href@noop {} {\emph {\bibinfo {title}
  {Multifractal volatility: theory, forecasting, and pricing}}}\ (\bibinfo
  {publisher} {Academic Press},\ \bibinfo {year} {2008})\BibitemShut {NoStop}%
\bibitem [{\citenamefont {Harte}(2001)}]{harte}%
  \BibitemOpen
  \bibfield  {author} {\bibinfo {author} {\bibfnamefont {David}\ \bibnamefont
  {Harte}},\ }\href@noop {} {\emph {\bibinfo {title} {Multifractals: theory and
  applications}}}\ (\bibinfo  {publisher} {CRC Press},\ \bibinfo {year}
  {2001})\BibitemShut {NoStop}%
\bibitem [{\citenamefont {Halsey}\ \emph {et~al.}(1986)\citenamefont {Halsey},
  \citenamefont {Jensen}, \citenamefont {Kadanoff}, \citenamefont {Procaccia},\
  and\ \citenamefont {Shraiman}}]{halsey1986fractal}%
  \BibitemOpen
  \bibfield  {author} {\bibinfo {author} {\bibfnamefont {Thomas~C}\
  \bibnamefont {Halsey}}, \bibinfo {author} {\bibfnamefont {Mogens~H}\
  \bibnamefont {Jensen}}, \bibinfo {author} {\bibfnamefont {Leo~P}\
  \bibnamefont {Kadanoff}}, \bibinfo {author} {\bibfnamefont {Itamar}\
  \bibnamefont {Procaccia}}, \ and\ \bibinfo {author} {\bibfnamefont {Boris~I}\
  \bibnamefont {Shraiman}},\ }\bibfield  {title} {\enquote {\bibinfo {title}
  {Fractal measures and their singularities: The characterization of strange
  sets},}\ }\href@noop {} {\bibfield  {journal} {\bibinfo  {journal} {Physical
  review A}\ }\textbf {\bibinfo {volume} {33}},\ \bibinfo {pages} {1141}
  (\bibinfo {year} {1986})}\BibitemShut {NoStop}%
\bibitem [{\citenamefont {Schertzer}\ \emph {et~al.}(1997)\citenamefont
  {Schertzer}, \citenamefont {Lovejoy}, \citenamefont {Schmitt}, \citenamefont
  {Chigirinskaya},\ and\ \citenamefont {Marsan}}]{schertzer1997multifractal}%
  \BibitemOpen
  \bibfield  {author} {\bibinfo {author} {\bibfnamefont {D}~\bibnamefont
  {Schertzer}}, \bibinfo {author} {\bibfnamefont {S}~\bibnamefont {Lovejoy}},
  \bibinfo {author} {\bibfnamefont {F}~\bibnamefont {Schmitt}}, \bibinfo
  {author} {\bibfnamefont {Y}~\bibnamefont {Chigirinskaya}}, \ and\ \bibinfo
  {author} {\bibfnamefont {D}~\bibnamefont {Marsan}},\ }\bibfield  {title}
  {\enquote {\bibinfo {title} {Multifractal cascade dynamics and turbulent
  intermittency},}\ }\href@noop {} {\bibfield  {journal} {\bibinfo  {journal}
  {Fractals}\ }\textbf {\bibinfo {volume} {5}},\ \bibinfo {pages} {427--471}
  (\bibinfo {year} {1997})}\BibitemShut {NoStop}%
\bibitem [{\citenamefont {Morales}\ \emph {et~al.}(2023)\citenamefont
  {Morales}, \citenamefont {Korbel},\ and\ \citenamefont {Rosas}}]{e25040678}%
  \BibitemOpen
  \bibfield  {author} {\bibinfo {author} {\bibfnamefont {Pablo~A.}\
  \bibnamefont {Morales}}, \bibinfo {author} {\bibfnamefont {Jan}\ \bibnamefont
  {Korbel}}, \ and\ \bibinfo {author} {\bibfnamefont {Fernando~E.}\
  \bibnamefont {Rosas}},\ }\bibfield  {title} {\enquote {\bibinfo {title}
  {Geometric structures induced by deformations of the legendre transform},}\
  }\href {\doibase 10.3390/e25040678} {\bibfield  {journal} {\bibinfo
  {journal} {Entropy}\ }\textbf {\bibinfo {volume} {25}} (\bibinfo {year}
  {2023}),\ 10.3390/e25040678}\BibitemShut {NoStop}%
\bibitem [{\citenamefont {Seifert}(2012)}]{seifert2012stochastic}%
  \BibitemOpen
  \bibfield  {author} {\bibinfo {author} {\bibfnamefont {Udo}\ \bibnamefont
  {Seifert}},\ }\bibfield  {title} {\enquote {\bibinfo {title} {Stochastic
  thermodynamics, fluctuation theorems and molecular machines},}\ }\href@noop
  {} {\bibfield  {journal} {\bibinfo  {journal} {Reports on progress in
  physics}\ }\textbf {\bibinfo {volume} {75}},\ \bibinfo {pages} {126001}
  (\bibinfo {year} {2012})}\BibitemShut {NoStop}%
\bibitem [{\citenamefont {Esposito}\ and\ \citenamefont {Van~den
  Broeck}(2010)}]{esposito2010three}%
  \BibitemOpen
  \bibfield  {author} {\bibinfo {author} {\bibfnamefont {Massimiliano}\
  \bibnamefont {Esposito}}\ and\ \bibinfo {author} {\bibfnamefont {Christian}\
  \bibnamefont {Van~den Broeck}},\ }\bibfield  {title} {\enquote {\bibinfo
  {title} {Three detailed fluctuation theorems},}\ }\href@noop {} {\bibfield
  {journal} {\bibinfo  {journal} {Physical review letters}\ }\textbf {\bibinfo
  {volume} {104}},\ \bibinfo {pages} {090601} (\bibinfo {year}
  {2010})}\BibitemShut {NoStop}%
\bibitem [{\citenamefont {Salamon}\ and\ \citenamefont
  {Berry}(1983)}]{salamon1983thermodynamic}%
  \BibitemOpen
  \bibfield  {author} {\bibinfo {author} {\bibfnamefont {Peter}\ \bibnamefont
  {Salamon}}\ and\ \bibinfo {author} {\bibfnamefont {R~Stephen}\ \bibnamefont
  {Berry}},\ }\bibfield  {title} {\enquote {\bibinfo {title} {Thermodynamic
  length and dissipated availability},}\ }\href@noop {} {\bibfield  {journal}
  {\bibinfo  {journal} {Physical Review Letters}\ }\textbf {\bibinfo {volume}
  {51}},\ \bibinfo {pages} {1127} (\bibinfo {year} {1983})}\BibitemShut
  {NoStop}%
\bibitem [{\citenamefont {Nulton}\ \emph {et~al.}(1985)\citenamefont {Nulton},
  \citenamefont {Salamon}, \citenamefont {Andresen},\ and\ \citenamefont
  {Anmin}}]{nulton1985quasistatic}%
  \BibitemOpen
  \bibfield  {author} {\bibinfo {author} {\bibfnamefont {J}~\bibnamefont
  {Nulton}}, \bibinfo {author} {\bibfnamefont {P}~\bibnamefont {Salamon}},
  \bibinfo {author} {\bibfnamefont {Bjarne}\ \bibnamefont {Andresen}}, \ and\
  \bibinfo {author} {\bibfnamefont {Qi}~\bibnamefont {Anmin}},\ }\bibfield
  {title} {\enquote {\bibinfo {title} {Quasistatic processes as step
  equilibrations},}\ }\href@noop {} {\bibfield  {journal} {\bibinfo  {journal}
  {The Journal of chemical physics}\ }\textbf {\bibinfo {volume} {83}},\
  \bibinfo {pages} {334--338} (\bibinfo {year} {1985})}\BibitemShut {NoStop}%
\bibitem [{\citenamefont {Bennett}(1976)}]{bennett1976efficient}%
  \BibitemOpen
  \bibfield  {author} {\bibinfo {author} {\bibfnamefont {Charles~H}\
  \bibnamefont {Bennett}},\ }\bibfield  {title} {\enquote {\bibinfo {title}
  {Efficient estimation of free energy differences from monte carlo data},}\
  }\href@noop {} {\bibfield  {journal} {\bibinfo  {journal} {Journal of
  Computational Physics}\ }\textbf {\bibinfo {volume} {22}},\ \bibinfo {pages}
  {245--268} (\bibinfo {year} {1976})}\BibitemShut {NoStop}%
\bibitem [{\citenamefont {Crooks}(2007)}]{crooks2007measuring}%
  \BibitemOpen
  \bibfield  {author} {\bibinfo {author} {\bibfnamefont {Gavin~E}\ \bibnamefont
  {Crooks}},\ }\bibfield  {title} {\enquote {\bibinfo {title} {Measuring
  thermodynamic length},}\ }\href@noop {} {\bibfield  {journal} {\bibinfo
  {journal} {Physical Review Letters}\ }\textbf {\bibinfo {volume} {99}},\
  \bibinfo {pages} {100602} (\bibinfo {year} {2007})}\BibitemShut {NoStop}%
\bibitem [{\citenamefont {Hanel}\ \emph {et~al.}(2012)\citenamefont {Hanel},
  \citenamefont {Thurner},\ and\ \citenamefont
  {Gell-Mann}}]{hanel2012generalized}%
  \BibitemOpen
  \bibfield  {author} {\bibinfo {author} {\bibfnamefont {Rudolf}\ \bibnamefont
  {Hanel}}, \bibinfo {author} {\bibfnamefont {Stefan}\ \bibnamefont {Thurner}},
  \ and\ \bibinfo {author} {\bibfnamefont {Murray}\ \bibnamefont {Gell-Mann}},\
  }\bibfield  {title} {\enquote {\bibinfo {title} {{Generalized entropies and
  logarithms and their duality relations}},}\ }\href@noop {} {\bibfield
  {journal} {\bibinfo  {journal} {Proceedings of the National Academy of
  Sciences}\ }\textbf {\bibinfo {volume} {109}},\ \bibinfo {pages}
  {19151--19154} (\bibinfo {year} {2012})}\BibitemShut {NoStop}%
\bibitem [{\citenamefont {Ducuara}\ \emph {et~al.}(2023)\citenamefont
  {Ducuara}, \citenamefont {Skrzypczyk}, \citenamefont {Buscemi}, \citenamefont
  {Sidajaya},\ and\ \citenamefont {Scarani}}]{ducuara2023maxwells}%
  \BibitemOpen
  \bibfield  {author} {\bibinfo {author} {\bibfnamefont {Andres~F.}\
  \bibnamefont {Ducuara}}, \bibinfo {author} {\bibfnamefont {Paul}\
  \bibnamefont {Skrzypczyk}}, \bibinfo {author} {\bibfnamefont {Francesco}\
  \bibnamefont {Buscemi}}, \bibinfo {author} {\bibfnamefont {Peter}\
  \bibnamefont {Sidajaya}}, \ and\ \bibinfo {author} {\bibfnamefont {Valerio}\
  \bibnamefont {Scarani}},\ }\href@noop {} {\enquote {\bibinfo {title}
  {Maxwell's demon walks into wall street: Stochastic thermodynamics meets
  expected utility theory},}\ } (\bibinfo {year} {2023}),\ \Eprint
  {http://arxiv.org/abs/2306.00449} {arXiv:2306.00449 [cond-mat.stat-mech]}
  \BibitemShut {NoStop}%
\bibitem [{\citenamefont {Korbel}(2021)}]{Korbel21a}%
  \BibitemOpen
  \bibfield  {author} {\bibinfo {author} {\bibfnamefont {Jan}\ \bibnamefont
  {Korbel}},\ }\bibfield  {title} {\enquote {\bibinfo {title} {Calibration
  invariance of the maxent distribution in the maximum entropy principle},}\
  }\href {\doibase 10.3390/e23010096} {\bibfield  {journal} {\bibinfo
  {journal} {Entropy}\ }\textbf {\bibinfo {volume} {23}} (\bibinfo {year}
  {2021}),\ 10.3390/e23010096}\BibitemShut {NoStop}%
\end{thebibliography}%
\end{document}